\documentclass[aps,pre,twocolumn,superscriptaddress,10pt]{revtex4-2}
\usepackage{graphicx,color}
\usepackage{amsmath,amsthm,amsfonts,amssymb,bm}
\usepackage{times}
\usepackage{epsf}
\usepackage[colorlinks=true]{hyperref}
\usepackage{mathrsfs}
\usepackage[T1]{fontenc}
\usepackage[utf8]{inputenc}
\usepackage{xcolor}
\usepackage{appendix}
\usepackage{algorithmic}
\usepackage{fullpage}
\usepackage{caption}

\hypersetup{
citecolor=blue,
filecolor=blue,
linkcolor=blue,
urlcolor=blue
}

\newcommand{\commentold}[1]{}
\DeclareMathSymbol{:}{\mathpunct}{operators}{"3A}

\usepackage{tikz}
\usetikzlibrary{shapes.geometric, arrows}

\tikzstyle{startstop} = [rectangle, rounded corners, minimum width=3cm, minimum height=1cm,text centered, draw=black, fill=gray!20]
\tikzstyle{process} = [rectangle, minimum width=3cm, minimum height=1cm, text centered, draw=black, fill=blue!10]
\tikzstyle{arrow} = [thick,->,>=stealth]

\begin{document}
\title{Optimizing the Charging of Open Quantum Batteries \\ using Long Short-Term Memory-Driven Reinforcement Learning}
\author{Shadab Zakavati}
\email{Sh.zakavati@uok.ac.ir \& shadab.zekavati@gmail.com}
\affiliation{Department of Physics, University of Kurdistan, P.O.Box 66177-15175, Sanandaj, Iran}
\author{Shahriar Salimi}
\affiliation{Department of Physics, University of Kurdistan, P.O.Box 66177-15175, Sanandaj, Iran}
\author{Behrouz Arash}
\email{behrouza@oslomet.no}
\affiliation{Department of Mechanical, Electrical, and Chemical Engineering, Oslo Metropolitan University, Pilestredet 35, 0166 Oslo, Norway}
\date{\today}

\begin{abstract}
Controlling the charging process of a quantum battery
 involves strategies to efficiently transfer, store, and retain energy, while mitigating decoherence, energy dissipation, and inefficiencies caused by surrounding interactions. We develop a model to study the charging process of a quantum battery in an open quantum setting, where the battery interacts with a charger and a structured reservoir. To overcome the limitations of static charging protocols, a reinforcement learning (RL) charging strategy is proposed, which utilizes the deep deterministic policy gradient algorithm alongside long short--term memory (LSTM) networks. The LSTM networks enable the RL model to capture temporal correlations driven by non-Markovian dynamics, facilitating a continuous, adaptive charging strategy. The RL protocols consistently outperform conventional fixed heuristic strategies by real-time controlling the driving field amplitude and coupling parameters. By penalizing battery-to-charger backflow in the reward function, the RL-optimized charging strategy promotes efficient unidirectional energy transfer from charger to battery, achieving higher and more stable extractable work. The proposed RL controller would provide a framework for designing efficient charging schemes in broader configurations and multi-cell quantum batteries.



\end{abstract}
\maketitle

\section{Introduction}
\label{intro}

Lately, there has been significant interest in exploring quantum thermodynamics, driven by the growing demand for device miniaturization \cite{Deffner2019, Binder2015a}. Examining thermodynamic principles within a quantum framework holds substantial value, both theoretically and practically \cite{Gemmer2009, Brandner2016, Goold2016}. A key goal of this expanding field of research is to develop various mechanisms and create devices capable of storing and transferring energy beyond the microscopic level, effectively acting as batteries. In this context, quantum batteries (QBs) are defined as finite-dimensional quantum systems designed to temporarily hold energy in quantum states and deliver it to other devices \cite{Alicki2013, Binder2015b, Deffner2017}.

QBs represent a transformative concept in quantum thermodynamics, with potential applications in quantum computing, nanoscale energy harvesting, quantum sensing \cite{Alicki2013, Campaioli2017}, spin systems \cite{Le2018b}, quantum cavities \cite{Fusco2016}, superconducting transmon qubits \cite{Santos2019}, Josephson quantum phase battery \cite{Strambini2020}, molecular battery \cite{Alicki2019}, Sachdev-Ye-Kitaev model \cite{Rossini2018}, and quantum oscillators \cite{Andolina2019b}. Unlike classical batteries, quantum batteries exploit entanglement, coherence, and superposition to achieve enhanced charging efficiency and work extraction \cite{Ferraro2018}. 

In much of the existing literature, QBs are treated as closed systems that undergo purely unitary evolution. However, given the delicate nature of quantum systems, it is reasonable to assume that these batteries may interact with their surrounding environment, resulting in the loss of stored energy. To address this challenge, the concept of open quantum batteries (OQBs) has emerged in recent years \cite{Santos2019, Santos2020, Rossini2019, Pirmoradian2019, Barra2019, Gherardini2020, Quach2020, Kamin2020, Farina2019, GPintos2020, zakavati2021bounds}. The dynamics of OQBs can be described using a set of completely positive and trace-preserving maps, which may exhibit either Markovian or non-Markovian behavior \cite{Breuer2002, Breuer2016, Breuer2009, zakavati2021bounds}. Interactions between OQBs and their reservoirs can cause energy dissipation and decoherence, making it critical to develop methods to stabilize energy storage and prevent energy leakage into the environment \cite{Gherardini2020, Quach2020, Kamin2020}. Recent studies have focused on minimizing energy dissipation in OQBs to mitigate these adverse effects \cite{Santos2019, Gherardini2020, Quach2020, Kamin2020}. While OQBs offer a realistic framework for practical applications, they also introduce complex dynamics that make optimization difficult \cite{Andolina2019}. Specifically, non-Markovian dynamics--marked by memory effects in system-environment interactions--present significant obstacles to optimizing charging power and extractable work, requiring innovative theoretical and computational methods to reconcile idealized models with experimental conditions \cite{Binder2015}.

The study of quantum batteries has progressed rapidly, with foundational contributions shaping the field. Alicki and Fannes \cite{Alicki2013} introduced the quantum battery concept, defining ergotropy as the maximum work extractable through unitary operations. Campaioli et al. \cite{Campaioli2017} demonstrated quantum advantages in collective charging protocols, leveraging entanglement to boost power output. For open systems, Ferraro et al. \cite{Ferraro2018} explored the impact of decoherence on solid-state OQBs, while Andolina et al. \cite{Andolina2019} analyzed dissipative charging in spin systems, highlighting the trade-off between speed and efficiency. These works, however, largely assume Markovian dynamics, limiting their applicability to realistic OQBs where non-Markovian effects dominate \cite{Binder2015}. Recent studies have begun addressing this gap. Santos et al. \cite{Santos2023} derived power bounds for non-Markovian quantum heat engines, suggesting memory can enhance energy transfer, but specific bounds for OQB charging remain elusive. Similarly, Manzano and Zambrini \cite{Manzano2024} investigated coherence-driven charging, revealing memory effects’ role in work extraction, yet without universal limits.

Optimization of OQBs' performance has also seen significant attention. Rossini et al. \cite{Rossini2019} employed gradient-based methods to optimize charging protocols, but these struggle with the high-dimensional parameter spaces of non-Markovian systems. Variational quantum algorithms, explored by Xu et al. \cite{Xu2021}, offer quantum-native solutions but scale poorly with system complexity. Machine learning, particularly reinforcement learning (RL), has emerged as a powerful tool for quantum control \cite{Bukov2018}. Bukov et al. \cite{Bukov2018} demonstrated RL’s efficacy in optimizing quantum gate sequences, while Zhang and Wang \cite{baba2023deep} applied deep RL to quantum thermal machines, achieving robust control under noise. An RL approach to optimize the charging process of a Dicke quantum battery demonstrates that adaptive control strategies can significantly enhance charging efficiency and energy storage capacity compared to traditional methods \cite{erdman2024reinforcement}. However, standard RL algorithms assume Markovian environments, faltering when memory effects are significant. Long Short-Term Memory (LSTM) networks, introduced by Hochreiter and Schmidhuber \cite{Hochreiter1997}, address temporal dependencies but have not been integrated into OQB optimization. Furthermore, energy backflow--an inefficiency where energy leaks from the battery to the charger--remains unaddressed, mainly in existing frameworks \cite{zakavati2021bounds}. Zakavati et al. \cite{zakavati2021bounds} quantified backflow in driven OQBs; however, optimization strategies to mitigate it are currently lacking. 

This work tackles these challenges with three key contributions. First, we provide a thermodynamic analysis of extractable work in OQBs within the Lindblad master equation framework, elucidating the interplay of coherence, dissipation, and entropy under non-Markovian dynamics. Second, we investigate the impact of energy backflow—where energy leaks from the battery to the charger on OQB performance, introducing a quantitative measure to assess its detrimental effects on charging efficiency. Third, we develop a novel RL framework integrating a Deep Deterministic Policy Gradient (DDPG) algorithm with LSTM networks, featuring a reward function that maximizes extractable work while minimizing backflow. This approach enables robust control of coupling parameter (\(\kappa\)) and driving field amplitude (\(\eta\)) across diverse environmental conditions (\(\gamma_1, \lambda, \Delta, T\)), with simulations demonstrating superior work extraction and efficiency compared to traditional methods, paving the way for scalable, high-performance OQB designs.

The paper is organized as follows. Section \ref{sec:OQBs} presents a general model for OQBs, including the Hamiltonian, dynamics, and thermodynamic quantities. Section \ref{sec:charging_protocol} illustrates a charging protocol for OQBs. Section \ref{sec:ml_model} details the RL framework, including the LSTM-based DDPG algorithm and backflow-aware reward function. Section \ref{sec:evaluation} evaluates performance through simulations, and Section \ref{sec:conclusion} discusses implications and future directions.

\section{Theoretical Framework for OQBs}
\label{sec:OQBs}

\begin{figure}[t]
	\includegraphics[scale=0.6]{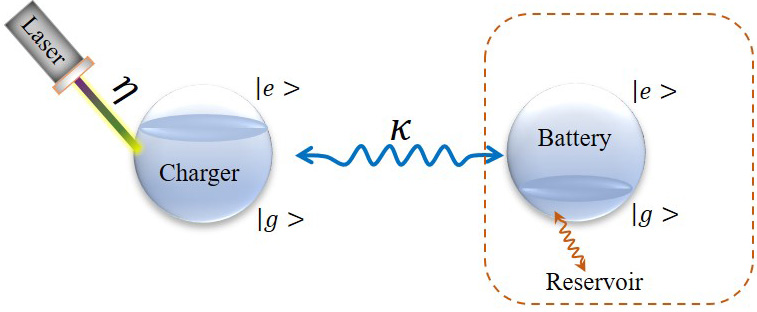}
	\caption{Schematic representation of an OQB system. The battery interacts with a charger during the charging process, while also being individually coupled to an environment. Additionally, an external field is applied to the charger.}
	\label{fig:OQB}
\end{figure}

We begin by introducing a general model for OQBs (see Fig.~\ref{fig:OQB}), comprising a quantum system acting as the battery, a charging mechanism, and interactions with a thermal bath within an open-system framework. The total Hamiltonian for the system, encompassing the charger ($A$), battery ($B$), and bath ($E$), is expressed as
\begin{eqnarray}\label{e11}
	H = H_A + H_B + H_E + H_{int},
\end{eqnarray}
where $H_A$, $H_B$, and $H_E$ represent the free Hamiltonians of the charger, battery, and bath, respectively, and $H_{int}$ encapsulates all interactions involving the quantum battery. Notably, $H_B$ is time-independent. In the interaction picture, the time evolution of the battery’s reduced density matrix at time $t$ is given by
\begin{eqnarray}\label{e12}
	\partial_t \rho = -i \, \text{Tr}_{(AE)}[H_{int}, \rho_{\text{tot}}],
\end{eqnarray}
where $\rho_{\text{tot}}$ denotes the density matrix of the entire system (battery, charger, and environment), and the partial trace is performed over the charger and environment subsystems. For a system coupled to a thermal bath, any non-equilibrium state possesses free energy that can be extracted as work. The non-equilibrium free energy is defined as
\begin{eqnarray}\label{e2}
	F(\rho) = U - \beta^{-1} S(\rho),
\end{eqnarray}
where $U = \text{Tr}(\rho H_B)$ is the system’s energy, $S(\rho) = -\text{Tr}(\rho \ln \rho)$ is the von Neumann entropy, and $\beta = T^{-1}$ is the inverse bath temperature, with Boltzmann’s constant set to $k_B = 1$ by convention \cite{Allahverdyan2004, Skrzypczyk2014, Brandao2013}. During relaxation, the system’s free energy decreases, reaching a minimum at the equilibrium state. Denoting the battery’s instantaneous state as $\rho$ and the thermal equilibrium state as $\tau_\beta = \frac{1}{Z} \exp(-\beta H_B)$, where $Z = \text{Tr}(\exp(-\beta H_B))$ is the partition function, the maximum extractable work is
\begin{equation}\label{e1}
	W_{\text{max}} = F(\rho) - F(\tau_\beta).
\end{equation}
Using the activity operator $\mathscr{A} := \beta^{-1} \ln \left( \frac{\rho}{\tau_\beta} \right)$ introduced in \cite{zakavati2021bounds}, the maximum extractable work can be reformulated as
\begin{equation}\label{e3}
	W_{\text{max}} = \frac{1}{\beta} \text{Tr} \left( \rho (\ln \rho - \ln \tau_\beta) \right) = \text{Tr}(\rho \, \mathscr{A}).
\end{equation}
At equilibrium, $\mathscr{A} = 0$, while $\mathscr{A} > 0$ for non-equilibrium states. The activity operator measures the system’s deviation from equilibrium, reflecting its capacity to yield extractable work. Notably, in \cite{GPintos2020}, the work operator $\mathbb{F} = H_B + \beta^{-1} \ln \rho$ has been proposed as an alternative formulation.

\renewcommand{\figurename}{FIG.}

\section{Quantum battery and dissipative/heating reservoir}
\label{sec:charging_protocol}

As previously outlined, the charging power and extractable work are defined using the activity operator, which depends on the inverse temperature $\beta$ and the system state $\rho$. The state of an open quantum system evolves according to the master equation, incorporating dissipation effects. Consequently, the system's state is highly sensitive to environmental conditions and interaction parameters that govern the master equation's solution. Thus, irrespective of the system's complexity and prior to examining specific examples, we anticipate that temperature and environmental parameters significantly influence the charging power and extractable work.

We now introduce a charging protocol where the quantum battery (QB) is coupled to a reservoir that induces dissipation and heating (see Fig.~\ref{fig2}). Consider a scenario where both the charger $A$ and the QB are two-qubit systems. The total Hamiltonian is given by \cite{Farina2019}
\begin{equation}\label{e38} 
	H=H_{0}+\Delta H_{A}+H_{AB}+H_{BE},
\end{equation}
where the first term is  the free Hamiltonian of the total system given by
\begin{equation} \label{e39}
	H_{0}=\frac{\omega_{0}}{2}(\sigma^{A}_{z}+1)+\frac{\omega_{0}}{2}(\sigma^{B}_{z}+1)+\sum_{k}\omega_{k}b^{\dagger}_{k}b_{k},
\end{equation}
and interaction Hamiltonians can be expressed as 
\begin{eqnarray} \label{e40}
	\Delta H_{A}=\eta (\sigma^{A}_{+} e^{- i\omega_{0}t}+\sigma^{A}_{-}e^{ i\omega_{0}t}),\nonumber\\
	H_{AB}=\kappa(\sigma^{A}_{+}\sigma^{B}_{-}+\sigma^{A}_{-}\sigma^{B}_{+}),\nonumber\\
	H_{BE}=\sum_{k}g_{k}(\sigma^{B}_{+}b_{k} +\sigma^{B}_{-}b^{\dagger}_{k}).
\end{eqnarray} 

In the above equation, $\sigma^{A,B}_{\pm}$ represent the raising and lowering operators for the respective qubits, $\omega_0$ and $\omega_k$ denote the transition frequencies of the qubits and the environment, respectively, $b_k$ ($b^\dagger_k$) are the annihilation (creation) operators for the $k$th mode of the bosonic environment, and $g_k$ is the coupling constant between the battery and the $k$th environmental mode. The first term in Eq.~(\ref{e40}), $\Delta H_A$, describes an external resonant driving field with amplitude $\eta$ that injects energy into the system, while the second term, $H_{AB}$, represents the interaction Hamiltonian between the charger and the battery, governed by the coupling parameter $\kappa$. The term $H_{BE}$ captures the interaction between the battery and the bath at temperature $T$. Notably, the charger does not interact with the bath.

In the interaction picture, the master equation for this model is explicitly given by \cite{Farina2019, Tabesh2018}
\begin{eqnarray}\label{e41}
	&&\frac{d\rho^{AB}}{dt}=-i[\kappa(\sigma^{A}_{+}\sigma^{B}_{-}+\sigma^{A}_{-}\sigma^{B}_{+})+\eta (\sigma^{A}_{+}+\sigma^{A}_{-}),\rho^{AB}]\nonumber\\
	&&+\dfrac{\gamma_{1}(t)}{2}(\sigma^{B}_{+}\rho^{AB}\sigma^{B}_{-}-\frac{1}{2}\{\sigma^{B}_{-}\sigma^{B}_{+},\rho^{AB}\})\nonumber\\
	&&+\dfrac{\gamma_{2}(t)}{2}(\sigma^{B}_{-}\rho^{AB}\sigma^{B}_{+}-\frac{1}{2}\{\sigma^{B}_{+}\sigma^{B}_{-},
	\rho^{AB}\}),
\end{eqnarray}
where $ \gamma_{1,2}$ shows time-dependent decay rates. The second and third terms describe heating and dissipation, respectively. 

Suppose the spectral density of the environment is taken as
\begin{eqnarray}\label{e44}
	J(\omega)= \gamma_0 \lambda^2 / 2\pi[(\omega_0 - \Delta - \omega)^2 + \lambda^2].
\end{eqnarray}

Here, $\gamma_0$ is an effective coupling constant linked to the battery system's relaxation time, $\tau_R \approx 1/\gamma_0$, while $\lambda$ characterizes the spectral width, related to the reservoir's correlation time, $\tau_B \approx 1/\lambda$. Additionally, $\Delta = \omega_0 - \nu_c$ represents the detuning, where $\nu_c$ is the central frequency of the thermal reservoir \cite{Breuer2002}. Such a spectrum effectively models an imperfect or leaky cavity. Accounting for these factors, the decay rates are expressed as $\gamma_1(t)/2 = N f(t)$ and $\gamma_2(t)/2 = (N+1) f(t)$, where $N = 1/[\exp(\omega_0/k_B T) - 1]$ denotes the mean photon number in the thermal reservoir's modes at temperature $T$, and $f(t)$ is determined by the reservoir's spectral density. At zero temperature, the heating rate vanishes ($\gamma_1(t) = 0$), and the dissipation rate simplifies to $\gamma_2(t)/2 = f(t)$ \cite{Breuer2002}. The function $f(t)$, derived in an exactly solvable form, is provided by \cite{Breuer2002}
\begin{eqnarray}\label{e42}
	&&f(t)=-2 \Re\{\frac{\dot{C}(t)}{C(t)}\},\nonumber\\
	&&C(t)= e^{-(\lambda-i\Delta)t/2}(\cosh(\frac{d t}{2})+\frac{\lambda-i\Delta}{d}\sinh(\frac{d t}{2}))C(0),\nonumber\\
\end{eqnarray}
where $d = \sqrt{(\lambda - i\Delta)^2 - 2\gamma_0 \lambda}$. We also define the ratio $R = \gamma_0 / \lambda$ to differentiate the strong coupling regime from the weak coupling regime. In the strong coupling regime, where $R \gg 1$, the function $f(t)$ may become negative during specific time intervals, leading to nondivisible, non-Markovian dynamics for the qubit \cite{Breuer2009, Breuer2012}.

In order to solve Eq. (\ref{e41}), we write $ \rho^{AB}$  in the matrix form
\begin{equation}\label{e43}
	\rho^{AB}(t)=\begin{pmatrix}
		\rho_{11}(t) &  \rho_{12}(t) &  \rho_{13}(t)  & \rho_{14}(t)  \\
		\rho_{21}(t) &  \rho_{22}(t) &  \rho_{23}(t)  & \rho_{24}(t) \\
		\rho_{31}(t) &  \rho_{32}(t) &  \rho_{33}(t)  & \rho_{34}(t) \\
		\rho_{41}(t) &  \rho_{42}(t) &  \rho_{43}(t)  & \rho_{44}(t) \\
	\end{pmatrix}.
\end{equation}

Substituting the above matrix into Eq. (\ref{e41}) gives a first-order system of ordinary differential equations in the sixteen unknown functions $\rho_{ij}(t)$, which has to be solved numerically under the initial conditions.

In OQBs, backflow refers to the unintended transfer of energy from the battery to the charger, reducing the efficiency of the charging process. This can occur due to non-unitary dynamics governed by the Lindblad master equation, where coherent interactions or environmental coupling lead to energy leakage from the battery's excited states back to the charger. To quantify this, we define the backflow as (see Appendix \ref{appendix A}):
\begin{equation}
	\mathscr{B} = \text{Im} \left( \text{tr} \left( \rho \sigma_A^+ \sigma_B^- \right) \right),
	\label{eq:backflow_def}
\end{equation}
where \(\rho\) is the density matrix of the quantum system, describing the combined state of the battery (subsystem B) and charger (subsystem A). \(\sigma_A^+\) represents the excited state of the battery, corresponding to a high-energy state where work is stored. \(\sigma_B^-\) represents the ground state of the charger, indicating a low-energy state. \(\sigma_A^+ \sigma_B^-\) is the outer product, a projection operator that captures transitions from the charger's ground state to the battery's excited state. 

The trace operation, \(\text{tr} \left( \rho \sigma_A^+ \sigma_B^- \right)\), computes the expectation value of this projector, reflecting the amplitude of the off-diagonal coherence between the battery's excited state and the charger's ground state. The imaginary part (\(\text{Im}\)) is extracted because it corresponds to the antisymmetric component of the dynamics, which is indicative of directional energy flow. Specifically:

\begin{itemize}
	\item A negative \(\text{backflow}\) value suggests that the system dynamics favor energy transfer from the battery back to the charger, reducing the stored work.
	\item A positive or zero \(\text{backflow}\) indicates that the charging process is proceeding without significant energy leakage, aligning with the goal of work maximization.
\end{itemize}

Physically, backflow arises from the interplay of coherent driving (controlled by \( \kappa \) and \( \eta \)) and dissipative effects (governed by environment parameters like \(\gamma_1\), \(\lambda\), \(\Delta\), and \(T\)). For instance, if the coupling parameter \( \kappa \) is too high, it may overdrive the system, causing oscillations that return energy to the charger. Similarly, a poorly tuned driving field amplitude \( \eta \) can exacerbate these effects in non-Markovian regimes, where memory effects amplify coherence-driven backflow. The backflow  definition in Eq.~\ref{eq:backflow_def} will later be used to design a reward function. The rationale for its effectiveness in maximizing work extraction will be detailed in Sec.~\ref{sec:ml_model}.

The maximum extractable can be obtained as:
\begin{equation}
	W_{max} = (\rho_{11} + \rho_{33}) - T S,
	\label{eq:work_def}
\end{equation}
where \( \rho_{11}, \rho_{33} \) are diagonal elements of the density matrix \(\rho\), representing populations in the battery’s energy eigenstates, and \( S \) is the von Neumann entropy. The term \( T S \) (where \( T \) is the temperature) accounts for the entropic cost of charging, ensuring that \( W_{max} \) represents the free energy available for work extraction.

\section{Machine Learning Approach for Maximizing Extractable Work}
\label{sec:ml_model}

In this section, we employ an RL approach to maximize the extractable work from OQBs. Traditional optimization techniques, such as analytical solutions or numerical methods, often struggle with the complexity of non-Markovian dynamics and the high-dimensional parameter spaces inherent in OQBs. RL, specifically the Deep Deterministic Policy Gradient (DDPG) algorithm augmented with LSTM networks, provides a robust framework to learn optimal control policies directly from interactions with the quantum environment, effectively handling memory effects and continuous control variables.

The RL agent optimizes the maximum extractable work (Eq.~\eqref{e3}) by learning a policy that maximizes the work, while accounting for dissipative and non-Markovian effects analyzed in Secs.~\ref{sec:OQBs} and \ref{sec:charging_protocol}. This section describes our RL model in detail. 

\subsection{Reinforcement Learning Framework}
\label{subsec:rl_framework}

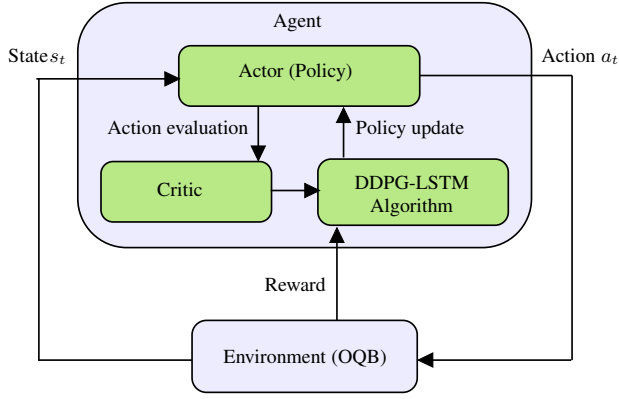
\begin{figure}[t]
	\centering
	\tikzset{every picture/.style={line width=0.75pt}} 
	\resizebox{0.5\textwidth}{!}{%
		\begin{tikzpicture}[x=0.75pt,y=0.75pt,yscale=-1,xscale=1]
			
			\draw  [fill={rgb, 255:red, 237; green, 236; blue, 254 }  ,fill opacity=1 ] (143.5,74.6) .. controls (143.5,58.25) and (156.75,45) .. (173.1,45) -- (387.9,45) .. controls (404.25,45) and (417.5,58.25) .. (417.5,74.6) -- (417.5,163.4) .. controls (417.5,179.75) and (404.25,193) .. (387.9,193) -- (173.1,193) .. controls (156.75,193) and (143.5,179.75) .. (143.5,163.4) -- cycle ;
			\draw  [color={rgb, 255:red, 0; green, 0; blue, 0 }  ,draw opacity=1 ][fill={rgb, 255:red, 184; green, 233; blue, 134 }  ,fill opacity=1 ] (204.5,78.3) .. controls (204.5,74.27) and (207.77,71) .. (211.8,71) -- (342.7,71) .. controls (346.73,71) and (350,74.27) .. (350,78.3) -- (350,100.2) .. controls (350,104.23) and (346.73,107.5) .. (342.7,107.5) -- (211.8,107.5) .. controls (207.77,107.5) and (204.5,104.23) .. (204.5,100.2) -- cycle ;
			\draw  [fill={rgb, 255:red, 184; green, 233; blue, 134 }  ,fill opacity=1 ] (288.5,147.5) .. controls (288.5,143.08) and (292.08,139.5) .. (296.5,139.5) -- (395.5,139.5) .. controls (399.92,139.5) and (403.5,143.08) .. (403.5,147.5) -- (403.5,171.5) .. controls (403.5,175.92) and (399.92,179.5) .. (395.5,179.5) -- (296.5,179.5) .. controls (292.08,179.5) and (288.5,175.92) .. (288.5,171.5) -- cycle ;
			\draw  [color={rgb, 255:red, 0; green, 0; blue, 0 }  ,draw opacity=1 ][fill={rgb, 255:red, 184; green, 233; blue, 134 }  ,fill opacity=1 ] (157.5,148.3) .. controls (157.5,144.27) and (160.77,141) .. (164.8,141) -- (253.2,141) .. controls (257.23,141) and (260.5,144.27) .. (260.5,148.3) -- (260.5,170.2) .. controls (260.5,174.23) and (257.23,177.5) .. (253.2,177.5) -- (164.8,177.5) .. controls (160.77,177.5) and (157.5,174.23) .. (157.5,170.2) -- cycle ;
			\draw  [fill={rgb, 255:red, 237; green, 236; blue, 254 }  ,fill opacity=1 ] (212.5,245.7) .. controls (212.5,240.9) and (216.4,237) .. (221.2,237) -- (339.8,237) .. controls (344.6,237) and (348.5,240.9) .. (348.5,245.7) -- (348.5,271.8) .. controls (348.5,276.6) and (344.6,280.5) .. (339.8,280.5) -- (221.2,280.5) .. controls (216.4,280.5) and (212.5,276.6) .. (212.5,271.8) -- cycle ;
			\draw    (118.5,91) -- (202,91) ;
			\draw [shift={(205,91)}, rotate = 179.67] [fill={rgb, 255:red, 0; green, 0; blue, 0 }  ][line width=0.08]  [draw opacity=0] (8.93,-4.29) -- (0,0) -- (8.93,4.29) -- cycle    ;
			\draw    (120,91) -- (120,261.5) ;
			\draw    (120,261) -- (212.5,261) ;
			\draw    (352,261) -- (441.5,261) ;
			\draw [shift={(349,261)}, rotate = 359.69] [fill={rgb, 255:red, 0; green, 0; blue, 0 }  ][line width=0.08]  [draw opacity=0] (8.93,-4.29) -- (0,0) -- (8.93,4.29) -- cycle    ;
			\draw    (350,89) -- (442,89) ;
			\draw    (442,89) -- (442,261) ;
			\draw    (300,237) -- (300,184) ;
			\draw [shift={(300,181)}, rotate = 90] [fill={rgb, 255:red, 0; green, 0; blue, 0 }  ][line width=0.08]  [draw opacity=0] (8.93,-4.29) -- (0,0) -- (8.93,4.29) -- cycle    ;
			\draw    (252.5,137.5) -- (252.5,107.5) ;
			\draw [shift={(252.5,140.5)}, rotate = 270] [fill={rgb, 255:red, 0; green, 0; blue, 0 }  ][line width=0.08]  [draw opacity=0] (8.93,-4.29) -- (0,0) -- (8.93,4.29) -- cycle    ;
			\draw    (260,158.5) -- (286.5,158.5) ;
			\draw [shift={(289.5,158.5)}, rotate = 180] [fill={rgb, 255:red, 0; green, 0; blue, 0 }  ][line width=0.08]  [draw opacity=0] (8.93,-4.29) -- (0,0) -- (8.93,4.29) -- cycle    ;
			\draw    (304,138.5) -- (304,111.5) ;
			\draw [shift={(304,108.5)}, rotate = 90] [fill={rgb, 255:red, 0; green, 0; blue, 0 }  ][line width=0.08]  [draw opacity=0] (8.93,-4.29) -- (0,0) -- (8.93,4.29) -- cycle    ;
			
			\draw (239,80.5) node [anchor=north west][inner sep=0.75pt]  [font=\small] [align=left] {Actor (Policy)};
			\draw (304,148) node [anchor=north west][inner sep=0.75pt]  [font=\small] [align=left] {\begin{minipage}[lt]{59.34pt}\setlength\topsep{0pt}
					\begin{center}
						DDPG-LSTM \\Algorithm
					\end{center}
					
			\end{minipage}};
			\draw (260,50) node [anchor=north west][inner sep=0.75pt]  [font=\small] [align=left] {Agent};
			\draw (190,152) node [anchor=north west][inner sep=0.75pt]  [font=\small] [align=left] {Critic};
			\draw (230,253) node [anchor=north west][inner sep=0.75pt]  [font=\small] [align=left] {Environment (OQB)};
			\draw (100,71) node [anchor=north west][inner sep=0.75pt]   [align=left] {State};
			\draw (125,74) node [anchor=north west][inner sep=0.75pt]    {$s_{t}$};
			\draw (422,71) node [anchor=north west][inner sep=0.75pt]   [align=left] {Action};
			\draw (458,74) node [anchor=north west][inner sep=0.75pt]    {$a_{t}$};
			\draw (255,209.5) node [anchor=north west][inner sep=0.75pt]   [align=left] {Reward};
			\draw (160,115) node [anchor=north west][inner sep=0.75pt]   [align=left] {{\small Action evaluation}};
			\draw (310,115) node [anchor=north west][inner sep=0.75pt]   [align=left] {Policy update};
		\end{tikzpicture}
	}
	\caption{Schematic of the DDPG-LSTM reinforcement learning framework for optimizing OQB.}	
	\label{fig:RLmodel}
\end{figure}

Our RL framework illustrated in Fig. \ref{fig:RLmodel} is built upon the DDPG algorithm, an off-policy actor-critic method designed for continuous action spaces~\cite{lillicrap2015continuous}. In DDPG, an \emph{actor} network \(\pi(s)\) maps states \(s\) to deterministic actions \(a\), while a \emph{critic} network \(Q(s,a)\) evaluates the expected cumulative discounted reward starting from state \(s\) when taking action \(a\). A key feature of DDPG is the use of \emph{target networks}—a target actor \(\pi'(s)\) and a target critic \(Q'(s,a)\)—which are slowly updated to stabilize training, a technique that mitigates divergence in Q-learning \cite{mnih2015human}. The environment is defined as follows:

\renewcommand{\figurename}{FIG.}

\begin{itemize}
	\item \textbf{State Space:} The state vector \(s\) is composed of two parts: 
	\begin{itemize}
		\item \emph{Battery State:} The state \(s\) is a 10-dimensional vector representing the quantum state of the battery, preprocessed into a 20-dimensional real vector by separating real and imaginary components.
		\item \emph{Environment State:} The environment is characterized by four parameters: temperature ($T$), $\lambda$, $\gamma_1$, and $\Delta$. These parameters are sampled uniformly within their respective physical ranges:
		\[
		T\in [0,1], \ \lambda\in [0.1, 1], \ \gamma_0\in [0.1, 1], \ \Delta\in [2,4].
		\]
	\end{itemize}
	The environmental parameters are normalized to the unit interval to ensure compatibility with the neural network input, a common practice in RL to improve training stability \cite{sutton2018reinforcement}. Thus, the overall state is constructed as the concatenation of the battery state (real and imaginary parts) and the normalized environment state.
	\item \textbf{Action Space:} The RL agent outputs a three-dimensional vector ($a=[a_1 \ a_2 \ a_3]$) which, after post-processing, yields two control signals:
	\begin{itemize}
		\item The first element is used to determine the control field $\kappa_{val}$ via a hyperbolic tangent transformation (i.e., $\eta_{val}=\text{tanh}(a_1)$), which is scaled into its physical interval $[\eta_{\min}, \eta_{\max}]$.
		\item The second and third elements are combined to set the coupling parameter $\kappa$. The raw value $\kappa_{\text{raw}}=\text{sig}(a_2)$ is passed through a sigmoid function (\(\text{sig}(x) = 1/(1 + e^{-x})\)), and a gating mechanism is applied: $\kappa_{val} = \text{gate}\cdot \kappa_{\text{raw}}$, where \(\text{gate} = \text{sig}(a_3)\). The gating mechanism enables controlling the backflow from the battery to the charger.
	\end{itemize}
	$\eta_{val}$ and $\kappa_{val}$ are scaled to the physical ranges via
	\begin{align}
		\eta = \eta_{\min} + 0.5(\eta_{\max} - \eta_{\min})(\eta_{\text{val}} + 1), \\ \kappa = \kappa_{\max} \cdot \kappa_{\text{val}},
		\label{eq:action_scaling2}
	\end{align}
	with \( \eta_{\min} = 0 \), \( \eta_{\max} = 100 \), \( \kappa_{\max} = 100 \).	 
	\item \textbf{Reward Function:} The reward \(r_t\) is designed to maximize the extractable work \(W_{\text{max}}\), potentially penalizing entropy production or energy inefficiencies, following principles of reward shaping in RL \cite{ng1999policy}.
\end{itemize}

The Q-value function is defined as:
\begin{equation}
	Q(s,a) = \mathbb{E}\left[\sum_{t=0}^{\infty} \gamma^t r_t \,\Big|\, s_0 = s, a_0 = a \right],
	\label{eq:qvalue_def}
\end{equation}
where \(\gamma \in (0,1)\) is the discount factor and \(r_t\) is the reward at time \(t\).

The target Q-value, used for training the critic, is computed using the target networks:
\begin{equation}
	y = r + \gamma\, Q'\big(s',\pi'(s')\big),
	\label{eq:target_Q}
\end{equation}
where \(s'\) is the state resulting from action \(a\) in state \(s\), and \(r\) is the immediate reward received at time \(t\) after taking action \(a\) in state \(s\).

In the reinforcement learning (RL) framework designed to optimize the charging of an OQB, the reward function \( r_t \) at time \( t \) is crafted to maximize the extractable work while ensuring stability in the control process. To optimize the charging process of open quantum batteries (OQBs), the reinforcement learning (RL) agent must maximize the extractable work \( W_{\text{max}} \) (Eq.~\eqref{e1}) while minimizing undesirable energy transfers from the battery back to the charger, a phenomenon we term \emph{backflow}. The reward function is carefully designed to balance these objectives, incorporating the extractable work and a penalty for backflow, enabling effective control of the system parameters \( \kappa \) (coupling parameter) and \( \eta \) (driving field amplitude) \cite{vinjanampathy2016quantum}. 

To incorporate backflow, the reward \( \rho_t \) at time step \( t \) is defined as:
\begin{equation}
	r_t = 
	\begin{cases} 
		W_{max} - B \cdot \text{gate}^2, & \text{if } \mathscr{B} < 0, \\
		W_{max}, & \text{otherwise},
	\end{cases}
	\label{eq:reward_backflow}
\end{equation}
where \( B = 1 \) is a penalty coefficient that scales the backflow penalty.

%

The \(\text{gate}^2\) penalty in Eq.~\eqref{eq:reward_backflow} ensures that the penalty is always positive and proportional to the actor’s confidence in the action, encouraging the agent to adjust \( \eta \) to minimize backflow \cite{ng1999policy}. 

Finally, the reward is normalized to the range \([0, 1]\) to stabilize training. This normalization ensures that the reward is well-scaled for the DDPG algorithm, preventing numerical issues during optimization \cite{mnih2015human}.

The proposed reward function effectively controls the parameters \( \kappa \) and \( \eta \) to maximize extractable work for several reasons:

\begin{itemize}
	\item \textbf{Direct Optimization of Work:} By setting the baseline reward to \( W_{max} \) (Eq.~\eqref{eq:work_def}), the agent is directly incentivized to increase the population in the battery’s high-energy states (\( \rho_{11}, \rho_{33} \)) while minimizing entropy \( S \). This aligns with the physical goal of storing maximal free energy in the battery, as \( W_{max} \) represents the work extractable under thermodynamic constraints \cite{vinjanampathy2016quantum}.
	
	\item \textbf{Backflow Penalty for Efficiency:} The conditional penalty \( -B \cdot \text{gate}^2 \) when \(\mathscr{B} < 0\) discourages actions that lead to energy leakage. Since backflow is tied to coherence effects that depend on \( \kappa \) and \( \eta \), the penalty guides the agent to select values that maintain forward energy transfer. The use of \(\text{gate}^2\) is particularly effective because it leverages the actor’s own output to modulate the penalty, allowing the agent to learn when to suppress \( \eta \) (via a low \(\text{gate}\)) to avoid backflow, thus fine-tuning the driving strength dynamically \cite{ng1999policy}.
	
	\item \textbf{Adaptability to Non-Markovian Dynamics:} The reward function operates within an LSTM-based DDPG framework, processing sequences of 10 time steps (see Sec.~\ref{subsec:net_arch_loss}). This allows the agent to capture temporal dependencies in the system’s evolution, such as memory effects that influence backflow \cite{heess2015memory}. By penalizing backflow over these sequences, the reward ensures that the agent learns policies robust to non-Markovian fluctuations, optimizing \( \kappa \) and \( \eta \) over time to sustain high work output.
	
	\item \textbf{Robustness Across Environments:} The training process samples environment parameters (\( T \in [0, 1] \), \(\lambda \in [0.1, 1] \), \(\gamma_1 \in [0.1, 1] \), \(\Delta \in [2, 4] \)) uniformly, embedding them in the state space. The reward’s structure—combining \( W \), which depends on \( T \), and a backflow penalty, which is environment-agnostic—ensures that the agent learns policies that generalize across varying dissipative conditions, maximizing work extraction regardless of external noise or detuning.
	
	\item \textbf{Numerical Stability:} The normalization maps the reward to \([0, 1]\), preventing large variations in \( W \) or the penalty term from destabilizing the critic’s Q-value estimates. This stability is crucial for convergence in DDPG, ensuring that the agent consistently improves its control over \( \kappa \) and \( \eta \) \cite{lillicrap2015continuous}.	
\end{itemize}

By balancing the drive to maximize \( W_{max} \) with the need to minimize backflow, the reward function enables the RL agent to explore and exploit the continuous action space effectively. For example, if the agent selects a high \( \kappa \) that increases \( W \) but triggers backflow, the penalty reduces the reward, prompting the agent to adjust \( \kappa \) or reduce \( \eta \) via the \(\text{gate}\) mechanism. Over time, the agent learns a policy that optimizes both parameters to achieve high work extraction with minimal energy loss, as demonstrated in related quantum control studies \cite{yao2023reinforcement}.. 

In summary, the reward function’s design—integrating extractable work, a physically motivated backflow penalty, and adaptive control via the \(\text{gate}\)—provides a robust mechanism for the RL agent to control the OQB’s charging process, achieving efficient and stable work maximization under complex quantum dynamics.

The RL model works as follows: the actor, critic, target actor, and target critic networks work in tandem. During training, the critic network is updated to minimize the error between its Q-value estimates and the target Q-values computed using the target networks. Simultaneously, the actor network is updated to maximize the expected Q-value estimated by the critic. Target networks are updated slowly via a soft update mechanism to maintain stability \cite{lillicrap2015continuous}. The overall training procedure is summarized in Algorithm 1. 

\subsection{Network Architectures and Loss Definitions}
\label{subsec:net_arch_loss}

The RL agent consists of an actor network and a critic network, designed to handle the continuous state and action spaces of the OQB environment \cite{lillicrap2015continuous}.

\paragraph{Actor Network.}  
The RL agent consists of an actor network and a critic network.

\paragraph{Actor Network.}
The actor processes a sequence input of dimension \(24 \times L\) through:
\begin{itemize}
	\item Two LSTM layers, each with 32 hidden units.
	\item Two fully connected layers with ReLU activations, reducing to a three-dimensional output.
	\item A final \(\tanh\) layer to produce actions in \([-1, 1]\).
\end{itemize}

\paragraph{Critic Network.}
The critic network has two branches:
\begin{itemize}
	\item \textbf{State Branch:} Two LSTM layers (32 hidden units each), followed by fully connected layers with ReLU activations.
	\item \textbf{Action Branch:} Two fully connected layers processing the three-dimensional action vector.
	\item Outputs are concatenated and passed through additional fully connected layers with ReLU activations to yield a scalar Q-value.
\end{itemize}

Mini-batches of sequences are sampled from the replay buffer. Let \(B\) denote the mini-batch size and \(L\) the sequence length. The data arrays have dimensions:
\begin{equation}
	S \in \mathbb{R}^{24 \times L \times B}, \quad A \in \mathbb{R}^{3 \times L \times B}.
	\label{eq:minibatch_dims}
\end{equation}

The critic loss is expressed in a double-series notation as
\begin{equation}
	L_{\text{critic}} = \frac{1}{B\,L} \sum_{i=1}^{B}\sum_{k=0}^{L-1} \left( Q\big(s_{i,k}, a_{i,k}\big) - y_{i,k} \right)^2,
	\label{eq:critic_loss_double}
\end{equation}
and the actor loss is written as
\begin{equation}
	L_{\text{actor}} = -\frac{1}{B\,L} \sum_{i=1}^{B}\sum_{k=0}^{L-1} Q\big(s_{i,k}, \pi(s_{i,k})\big).
	\label{eq:actor_loss_double}
\end{equation}
These loss functions follow the standard DDPG formulation, ensuring that the critic learns accurate Q-values while the actor improves the policy \cite{lillicrap2015continuous}.

\subsection{Handling Non-Markovian Dynamics with LSTM Networks}
\label{subsec:lstm_nonmarkov}

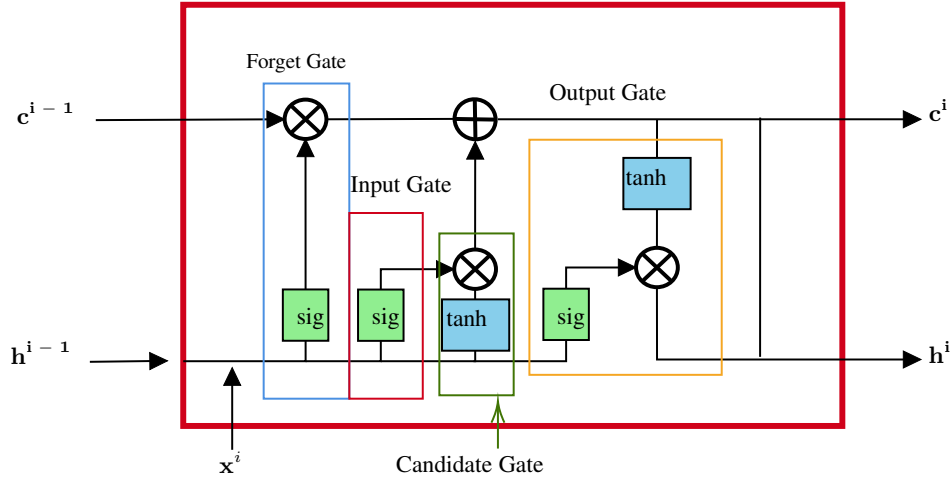
\begin{figure*}[!ht]
	\centering	
	\tikzset{every picture/.style={line width=0.75pt}} 
	\begin{tikzpicture}[x=0.75pt,y=0.75pt,yscale=-1,xscale=1]
		\draw  [color={rgb, 255:red, 208; green, 2; blue, 27 }  ,draw opacity=0.8 ][line width=2.25]  (174.1,22) -- (506.51,22) -- (506.51,234.59) -- (174.1,234.59) -- cycle ;
		\draw    (127,202.22) -- (163.16,201.76) ;
		\draw [shift={(166.16,201.72)}, rotate = 179.28] [fill={rgb, 255:red, 0; green, 0; blue, 0 }  ][line width=0.08]  [draw opacity=0] (8.93,-4.29) -- (0,0) -- (8.93,4.29) -- cycle    ;
		\draw    (174.1,201.91) -- (235.82,201.91) -- (235.77,92.74) ;
		\draw [shift={(235.77,89.74)}, rotate = 89.98] [fill={rgb, 255:red, 0; green, 0; blue, 0 }  ][line width=0.08]  [draw opacity=0] (8.93,-4.29) -- (0,0) -- (8.93,4.29) -- cycle    ;
		\draw    (130,80.22) -- (222.24,79.81) ;
		\draw [shift={(225.24,79.8)}, rotate = 179.75] [fill={rgb, 255:red, 0; green, 0; blue, 0 }  ][line width=0.08]  [draw opacity=0] (8.93,-4.29) -- (0,0) -- (8.93,4.29) -- cycle    ;
		\draw  [line width=1.5]  (225.24,79.8) .. controls (225.24,74.31) and (229.96,69.87) .. (235.77,69.87) .. controls (241.59,69.87) and (246.31,74.31) .. (246.31,79.8) .. controls (246.31,85.29) and (241.59,89.74) .. (235.77,89.74) .. controls (229.96,89.74) and (225.24,85.29) .. (225.24,79.8) -- cycle ; 
		\draw  [line width=1.5]  (228.33,72.78) -- (243.22,86.83) ; 
		\draw  [line width=1.5]  (243.22,72.78) -- (228.33,86.83) ;
		\draw    (246.31,79.8) -- (310.67,79.61) ;
		\draw  [line width=1.5]  (313.85,72.54) .. controls (317.83,68.66) and (324.39,68.76) .. (328.51,72.78) .. controls (332.62,76.79) and (332.73,83.18) .. (328.75,87.06) .. controls (324.77,90.94) and (318.21,90.84) .. (314.1,86.83) .. controls (309.98,82.81) and (309.87,76.42) .. (313.85,72.54) -- cycle ; 
		\draw  [line width=1.5]  (321.13,69.7) -- (321.47,89.9) ; 
		\draw  [line width=1.5]  (331.66,79.97) -- (310.94,79.63) ;
		\draw    (332.72,79.8) -- (543.19,79.8) ;
		\draw [shift={(546.19,79.8)}, rotate = 180] [fill={rgb, 255:red, 0; green, 0; blue, 0 }  ][line width=0.08]  [draw opacity=0] (8.93,-4.29) -- (0,0) -- (8.93,4.29) -- cycle    ;
		\draw    (235.82,201.91) -- (273.74,201.91) -- (273.74,155.47) -- (307.77,155.47) ;
		\draw [shift={(310.77,155.47)}, rotate = 180] [fill={rgb, 255:red, 0; green, 0; blue, 0 }  ][line width=0.08]  [draw opacity=0] (8.93,-4.29) -- (0,0) -- (8.93,4.29) -- cycle    ;
		\draw    (273.74,201.91) -- (321.35,201.91) -- (321.3,165.41) ;
		\draw  [line width=1.5]  (310.77,155.47) .. controls (310.77,149.99) and (315.48,145.54) .. (321.3,145.54) .. controls (327.12,145.54) and (331.83,149.99) .. (331.83,155.47) .. controls (331.83,160.96) and (327.12,165.41) .. (321.3,165.41) .. controls (315.48,165.41) and (310.77,160.96) .. (310.77,155.47) -- cycle ; 
		\draw  [line width=1.5]  (313.85,148.45) -- (328.75,162.5) ; 
		\draw  [line width=1.5]  (328.75,148.45) -- (313.85,162.5) ;
		\draw    (320.47,201.91) -- (367.2,201.91) -- (367.2,154.61) -- (399.47,154.61) ;
		\draw [shift={(402.47,154.61)}, rotate = 180] [fill={rgb, 255:red, 0; green, 0; blue, 0 }  ][line width=0.08]  [draw opacity=0] (8.93,-4.29) -- (0,0) -- (8.93,4.29) -- cycle    ;
		\draw    (321.3,145.54) -- (321.32,121.94) -- (321.35,93.98) ;
		\draw [shift={(321.35,90.98)}, rotate = 90.05] [fill={rgb, 255:red, 0; green, 0; blue, 0 }  ][line width=0.08]  [draw opacity=0] (8.93,-4.29) -- (0,0) -- (8.93,4.29) -- cycle    ;
		\draw  [line width=1.5]  (402.47,154.61) .. controls (402.47,149.13) and (407.19,144.68) .. (413,144.68) .. controls (418.82,144.68) and (423.54,149.13) .. (423.54,154.61) .. controls (423.54,160.1) and (418.82,164.55) .. (413,164.55) .. controls (407.19,164.55) and (402.47,160.1) .. (402.47,154.61) -- cycle ; 
		\draw  [line width=1.5]  (405.55,147.59) -- (420.45,161.64) ; 
		\draw  [line width=1.5]  (420.45,147.59) -- (405.55,161.64) ;
		\draw    (413,144.68) -- (413.05,79.8) ;
		\draw    (413,164.55) -- (413.05,201.05) -- (543.19,201.05) ;
		\draw [shift={(546.19,201.05)}, rotate = 180] [fill={rgb, 255:red, 0; green, 0; blue, 0 }  ][line width=0.08]  [draw opacity=0] (8.93,-4.29) -- (0,0) -- (8.93,4.29) -- cycle    ;
		\draw    (465.03,199.71) -- (465.07,78.94) ;
		\draw  [fill={rgb, 255:red, 144; green, 238; blue, 144 }  ,fill opacity=1 ] (224.33,165.69) -- (247.28,165.69) -- (247.28,191.59) -- (224.33,191.59) -- cycle ;
		\draw  [fill={rgb, 255:red, 144; green, 238; blue, 144 }  ,fill opacity=1 ] (262.26,165.74) -- (285.21,165.74) -- (285.21,191.64) -- (262.26,191.64) -- cycle ;
		\draw  [fill={rgb, 255:red, 135; green, 206; blue, 235 }  ,fill opacity=1 ] (304.6,170.71) -- (338.52,170.71) -- (338.52,196.61) -- (304.6,196.61) -- cycle ;
		\draw  [fill={rgb, 255:red, 144; green, 238; blue, 144 }  ,fill opacity=1 ] (355.73,165.31) -- (378.67,165.31) -- (378.67,191.21) -- (355.73,191.21) -- cycle ;
		\draw  [fill={rgb, 255:red, 135; green, 206; blue, 235 }  ,fill opacity=1 ] (396.07,99.29) -- (429.99,99.29) -- (429.99,125.19) -- (396.07,125.19) -- cycle ;
		\draw    (198.79,246.62) -- (198.79,208.16) ;
		\draw [shift={(198.79,205.16)}, rotate = 90] [fill={rgb, 255:red, 0; green, 0; blue, 0 }  ][line width=0.08]  [draw opacity=0] (8.93,-4.29) -- (0,0) -- (8.93,4.29) -- cycle    ;
		\draw  [color={rgb, 255:red, 74; green, 144; blue, 226 }  ,draw opacity=1 ] (214.66,61.74) -- (257.86,61.74) -- (257.86,220.83) -- (214.66,220.83) -- cycle ;
		\draw  [color={rgb, 255:red, 208; green, 2; blue, 27 }  ,draw opacity=1 ] (257.86,127.1) -- (294.9,127.1) -- (294.9,220.83) -- (257.86,220.83) -- cycle ;
		\draw  [color={rgb, 255:red, 245; green, 166; blue, 35 }  ,draw opacity=1 ] (348.68,90.12) -- (445.67,90.12) -- (445.67,208.79) -- (348.68,208.79) -- cycle ;
		\draw  [color={rgb, 255:red, 65; green, 117; blue, 5 }  ,draw opacity=1 ] (303.27,137.42) -- (340.75,137.42) -- (340.75,219.11) -- (303.27,219.11) -- cycle ;
		%
		\draw [color={rgb, 255:red, 65; green, 117; blue, 5 }  ,draw opacity=1 ]   (332.81,246) -- (332.81,225) ;
		\draw [shift={(332.81,222)}, rotate = 90] [color={rgb, 255:red, 65; green, 117; blue, 5 }  ,draw opacity=1 ][line width=0.75]    (10.93,-3.29) .. controls (6.95,-1.4) and (3.31,-0.3) .. (0,0) .. controls (3.31,0.3) and (6.95,1.4) .. (10.93,3.29);		
		%
		\draw (229.82,173.51) node [anchor=north west][inner sep=0.75pt]   [align=left] {$\displaystyle \mathbf{\text{sig} }$};
		\draw (267.74,173.51) node [anchor=north west][inner sep=0.75pt]   [align=left] {$\displaystyle \mathbf{\text{sig} }$};
		\draw (305.29,174.66) node [anchor=north west][inner sep=0.75pt]   [align=left] {$\displaystyle \text{tanh}$};
		\draw (361.2,173.51) node [anchor=north west][inner sep=0.75pt]   [align=left] {$\displaystyle \mathbf{\text{sig} }$};
		\draw (395.88,103.19) node [anchor=north west][inner sep=0.75pt]   [align=left] {$\displaystyle \text{tanh}$};
		\draw (88.69,69.61) node [anchor=north west][inner sep=0.75pt]   [align=left] {$\displaystyle \mathbf{c^{i\ -\ 1}}$};
		\draw (548.6,67.89) node [anchor=north west][inner sep=0.75pt]   [align=left] {$\displaystyle \mathbf{c^{i}}$};
		\draw (548.42,190) node [anchor=north west][inner sep=0.75pt]   [align=left] {$\displaystyle \mathbf{h^{i}}$};
		\draw (85.45,190.28) node [anchor=north west][inner sep=0.75pt]   [align=left] {$\displaystyle \mathbf{h^{i\ -\ 1}}$};
		\draw (190.73,245.9) node [anchor=north west][inner sep=0.75pt]   [align=left] {$\displaystyle \mathbf{x}^{i}$};
		\draw (204.58,45.17) node [anchor=north west][inner sep=0.75pt]  [font=\footnotesize] [align=left] {Forget Gate};
		\draw (257,107.08) node [anchor=north west][inner sep=0.75pt]  [font=\small] [align=left] {Input Gate};
		\draw (357.3,60.51) node [anchor=north west][inner sep=0.75pt]  [align=left] {Output Gate};
		\draw (280,248) node [anchor=north west][inner sep=0.75pt]  [align=left] {Candidate Gate};
	\end{tikzpicture}
	\caption{A single LSTM cell. Here, green boxes represent the sigmoid ($\text{sig}$) activation, and blue boxes represent the hyperbolic tangent ($\tanh$) activation.}
	\label{fig: lstm}
\end{figure*}

Due to the inherent non-Markovian dynamics in quantum battery systems, it is essential to capture temporal dependencies across multiple time steps. LSTM networks achieve this by maintaining an internal cell state \(c_t\) that is updated over time \cite{hochreiter1997long}. A standard LSTM cell operates via the following equations:
\begin{align}
	f_t &= \sigma\left(W_f x_t + U_f h_{t-1} + b_f\right), \label{eq:lstm_forget}\\[1mm]
	i_t &= \sigma\left(W_i x_t + U_i h_{t-1} + b_i\right), \label{eq:lstm_input}\\[1mm]
	\tilde{c}_t &= \tanh\left(W_c x_t + U_c h_{t-1} + b_c\right), \label{eq:lstm_candidate}\\[1mm]
	c_t &= f_t \odot c_{t-1} + i_t \odot \tilde{c}_t, \label{eq:lstm_cell}\\[1mm]
	o_t &= \sigma\left(W_o x_t + U_o h_{t-1} + b_o\right), \label{eq:lstm_output}\\[1mm]
	h_t &= o_t \odot \tanh(c_t), \label{eq:lstm_hidden}
\end{align}
where \(x_t\) is the input at time \(t\), \(h_t\) is the hidden state, and \(c_t\) is the cell state. As presented in Fig.~\ref{fig: lstm}, \(f_t\), \(i_t\), and \(o_t\) are the forget, input, and output gates, respectively. Also, \(W\), \(U\), and \(b\) are respectively weight matrices and bias vectors. \(\sigma(\cdot)\) is the sigmoid function, and \(\odot\) denotes element-wise multiplication. These equations enable the LSTM to learn long-term dependencies critical for modeling memory effects (non-Markovian dynamics) in our quantum battery environment \cite{heess2015memory}.

\vspace{2mm}
\noindent\textbf{Hyperparameters for the LSTM Networks.}  
Table~\ref{tab:hyperparams} lists the key hyperparameters used in our model, selected based on empirical tuning and common practices in deep RL \cite{lillicrap2015continuous}.

\begin{table}[ht]
	\centering
	\caption{Hyperparameters for the LSTM Networks}
	\begin{tabular}{l l l}
		\hline
		\textbf{Hyperparameter} & \textbf{Symbol} & \textbf{Value} \\
		\hline
		Number of LSTM Layers & \(n_{\text{LSTM}}\) & 2 \\
		Number of Hidden Units per Layer & \(H\) & 32 \\
		Mini-batch Sequence Length & \(L\) & 10 \\
		Mini-batch Size & \(B\) & 256 \\
		Learning Rate (Actor) & \(\eta_{\text{actor}}\) & \(1 \times 10^{-4}\) \\
		Learning Rate (Critic) & \(\eta_{\text{critic}}\) & \(1 \times 10^{-3}\) \\
		Discount Factor & \(\gamma\) & 0.99 \\
		Soft Update Coefficient & \(\tau\) & 0.005 \\
		Exploration Noise Std. Dev. & \(\sigma\) & 0.1 \\
		\hline
	\end{tabular}
	\label{tab:hyperparams}
\end{table}

\subsection{Network Updating Procedure}
\label{subsec:network_update}

After generating data in parallel (see Algorithm~3), mini-batches of sequences are sampled from the replay buffer \cite{lin1992self}. A sampled mini-batch consists of sequences
\begin{equation}
	\{(s_t^{(i)},a_t^{(i)},r_t^{(i)},s_{t+1}^{(i)},d_t^{(i)})\}_{t=t_0}^{t_0+L-1}, \quad i=1,\ldots,B,
	\label{eq:minibatch_sampling}
\end{equation}
and the discounted return for each sequence is computed as
\begin{equation}
	G_t = \sum_{k=0}^{L-1} \gamma^k r_{t+k}.
	\label{eq:discounted_return}
\end{equation}
The network parameters are updated using gradient descent:
\begin{equation}
	\theta \leftarrow \theta - \eta\,\nabla_\theta L,
	\label{eq:parameter_update}
\end{equation}
where \(\eta\) is the learning rate. Target networks are updated via a soft update:
\begin{equation}
	\theta' \leftarrow \tau\,\theta + (1-\tau)\,\theta',
	\label{eq:soft_update}
\end{equation}
with \(\tau=0.005\), a method that improves stability in DDPG training \cite{lillicrap2015continuous}. Algorithm~2 details the soft update procedure.

Algorithm~3 outlines the overall RL model training procedure, showing how the actor, critic, target actor, and target critic networks interact to update the actor network, following the DDPG methodology \cite{lillicrap2015continuous}.

\begin{figure*}[ht]
	\centering
	\label{alg:rl_training}
	\caption*{Algorithm 1. RL Model Training Procedure}
	\hrule height 1pt
	\vspace{2mm}
	\begin{algorithmic}[1]
		\STATE \textbf{Input:} Global replay buffer \(\mathcal{B}\), number of training updates \(N_{\text{update}}\)
		\FOR{\(u = 1\) to \(N_{\text{update}}\)}
		\STATE Sample a mini-batch of sequences \(\{(s_{i,k}, a_{i,k}, r_{i,k}, s'_{i,k}, d_{i,k})\}\) with \(i=1,\ldots,B\) and \(k=0,\ldots,L-1\)
		\STATE Compute target Q-values: 
		\[
		y_{i,k} = r_{i,k} + \gamma\, Q'\big(s'_{i,k}, \pi'(s'_{i,k})\big)
		\]
		\STATE Update critic network by minimizing
		\[
		L_{\text{critic}} = \frac{1}{B\,L} \sum_{i=1}^{B}\sum_{k=0}^{L-1} \left( Q(s_{i,k}, a_{i,k}) - y_{i,k} \right)^2
		\]
		\STATE Update actor network by minimizing
		\[
		L_{\text{actor}} = -\frac{1}{B\,L} \sum_{i=1}^{B}\sum_{k=0}^{L-1} Q\big(s_{i,k}, \pi(s_{i,k})\big)
		\]
		\STATE Soft update target networks using Algorithm~2 
		\ENDFOR
		\STATE \textbf{Output:} Updated actor and critic networks
	\end{algorithmic}
	\vspace{2mm}
	\hrule height 1pt
\end{figure*}

\begin{figure*}[ht]
	\centering
	\caption*{Algorithm 2. Soft Update for Target Networks}
	\hrule height 1pt
	\vspace{2mm}
	\begin{algorithmic}[1]
		\STATE \textbf{Input:} Main network parameters \(\theta\), target network parameters \(\theta'\), soft update coefficient \(\tau\)
		\FORALL{learnable parameters \(i\)}
		\STATE \(\theta'_i \gets \tau\,\theta_i + (1-\tau)\,\theta'_i\)
		\ENDFOR
		\STATE \textbf{Output:} Updated target network parameters \(\theta'\)
	\end{algorithmic}
	\vspace{2mm}
	\hrule height 1pt
	\label{alg:soft_update}
\end{figure*}

\begin{figure*}[ht]
	\centering
	\caption*{Algorithm 3. Parallel Data Generation for RL Training}
	\hrule height 1pt
	\vspace{2mm}
	\begin{algorithmic}[1]
		\STATE \textbf{Input:} Total episodes \(N_{\text{epi}}\), maximum steps per episode \(N_{\text{steps}}\), environment parameters, buffer capacity, sequence length \(L\)
		\FORALL{episode \(=1\) to \(N_{\text{epi}}\) in parallel}
		\STATE Initialize local replay buffer \(\mathcal{B}_{\text{local}}\)
		\STATE Reset environment: \(s \gets \text{resetEnv()}\)
		\FOR{\(t=1\) to \(N_{\text{steps}}\)}
		\STATE Preprocess \(s\) into real and imaginary parts.
		\STATE \(a \gets \pi(s)\) with added Gaussian noise (\(\sigma=0.2\))
		\STATE Scale \(a\) using Eq.~\eqref{eq:action_scaling}
		\STATE Execute \(a\); obtain next state \(s'\) and reward \(r\)
		\STATE Store \((s,a,r,s',d)\) in \(\mathcal{B}_{\text{local}}\)
		\STATE \(s \gets s'\)
		\IF{terminal condition \(d\) is met}
		\STATE \textbf{break}
		\ENDIF
		\ENDFOR
		\STATE Optionally refine rewards in \(\mathcal{B}_{\text{local}}\)
		\STATE Save \(\mathcal{B}_{\text{local}}\) to the global replay buffer \(\mathcal{B}\)
		\ENDFOR
		\STATE \textbf{Output:} Global replay buffer \(\mathcal{B}\)
	\end{algorithmic}
	\vspace{2mm}
	\hrule height 1pt
	\label{alg:parallel_data}
\end{figure*}

\section{Discussion}
\label{sec:evaluation}

In this work, we have introduced a unified framework that merges rigorous analytical power bounds with a RL strategy to optimize the charging of OQBs under realistic, non--Markovian dynamics. Our approach leverages LSTM--augmented DDPG to learn feedback‐driven control policies for the coupling parameter $\kappa$ and driving field amplitude $\eta$, while respecting the fundamental limits set by the Lindblad master equation. This section interprets the results, focusing on paired plots of maximum extractable work (\(W_{max}\)) and Coupling parameter (\(\kappa\)) versus time for each parameter set, emphasizing the RL model’s control of battery-charger interactions to prevent backflow and the role of environment-to-battery backflow in inducing non-Markovian dynamics.

\begin{figure}[t]
	\includegraphics[scale=0.5]{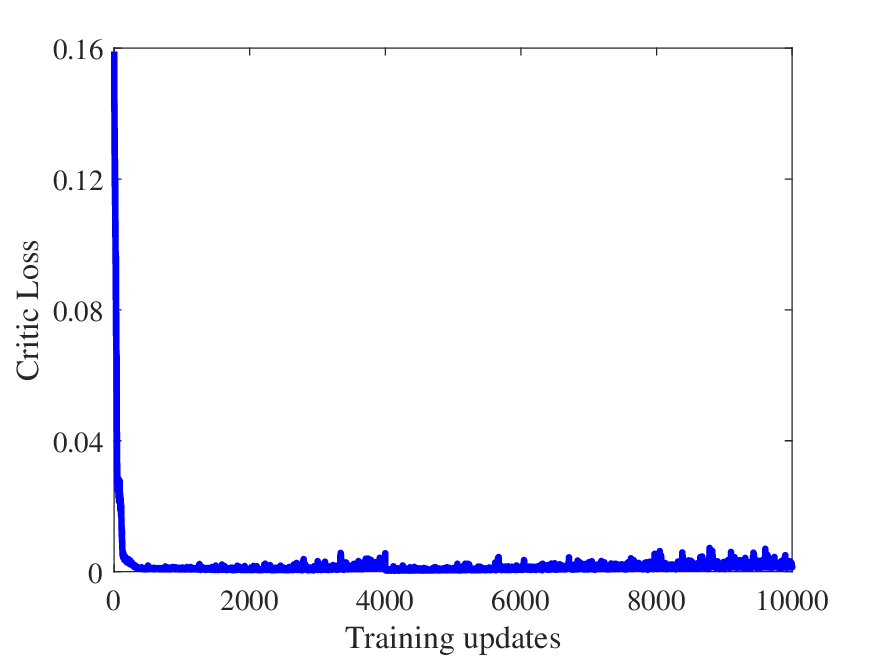}
	\caption{Training performance of the DDPG-LSTM model over 10000 updates. The critic loss shows rapid convergence to near zero with minor spikes.}
	\label{fig:critic_loss}
\end{figure}

\begin{figure}[t]
	\includegraphics[scale=0.5]{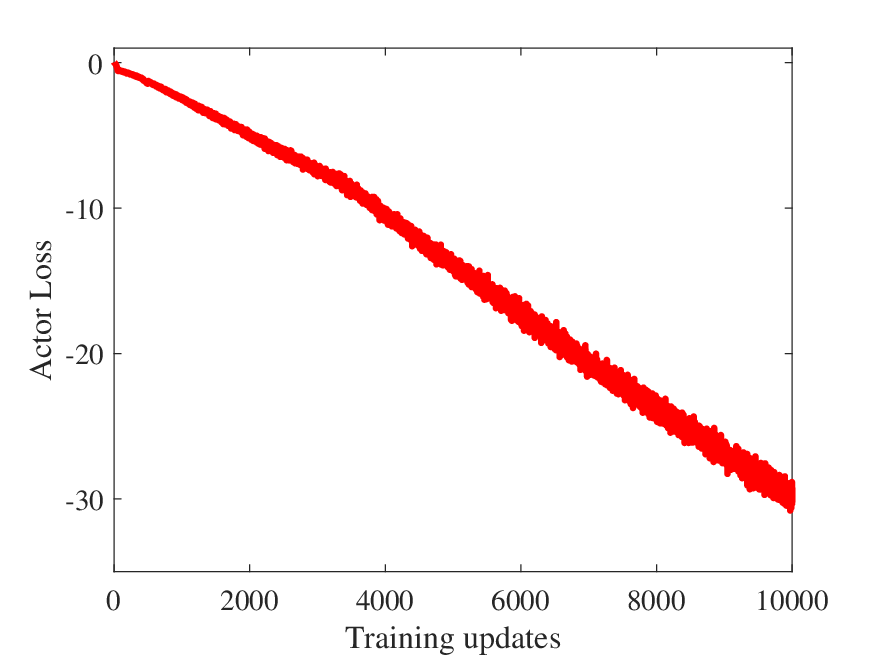}
	\caption{Training performance of the DDPG-LSTM model over 10000 updates. The actor loss decreases steadily, indicating improved policy performance.}
	\label{fig:actor_loss}
\end{figure}

The training performance of the DDPG-LSTM model is illustrated in Figs.~\ref{fig:critic_loss} and \ref{fig:actor_loss}, showing the critic loss and actor loss over 10000 updates. In Fig.~\ref{fig:critic_loss}, the critic loss, measured as the mean squared temporal difference error, starts at approximately 0.15 and decreases sharply within the first 1000 updates, dropping to near zero. After this initial drop, the loss remains low, fluctuating slightly around zero with minor spikes (e.g., around updates 5000 and 8000), and stabilizes close to zero by 10000 updates. This sharp initial decrease indicates rapid convergence of the critic network, which learns to accurately estimate the Q-values for state-action pairs in the OQB environment. The low and stable loss suggests reliable Q-value predictions, essential for guiding the actor’s policy updates in the DDPG framework. The minor spikes likely result from exploration noise in the actor’s actions or updates to the target critic network, introducing temporary TD errors.

In Fig.~\ref{fig:actor_loss}(b), the actor’s loss, defined as the negative Q-value plus a backflow penalty (\( B \cdot \text{gate}^2 \), with \( B = 1 \)), begins at zero and decreases steadily over the 10000 updates, reaching approximately -30 by the end, with some noise throughout. The actor’s objective is to maximize the Q-value by selecting optimal actions \( \kappa \) and \( \eta \), controlling the coupling strength and driving field amplitude of the OQB. The decreasing loss indicates that the actor learns a policy that improves the expected future reward (i.e., extractable work), with a more negative loss corresponding to a higher Q-value, reflecting enhanced work extraction. The backflow penalty ensures the policy minimizes energy backflow, and its influence is evident in the downward trend, suggesting a reduction in backflow alongside Q-value optimization. The noise in the loss curve arises from exploration in the DDPG algorithm and the non-Markovian dynamics of the OQB environment, which the LSTM component addresses by capturing temporal dependencies.

\renewcommand{\figurename}{FIG.}

\renewcommand{\figurename}{FIG.}

\begin{figure}[t]
	\includegraphics[scale=0.5]{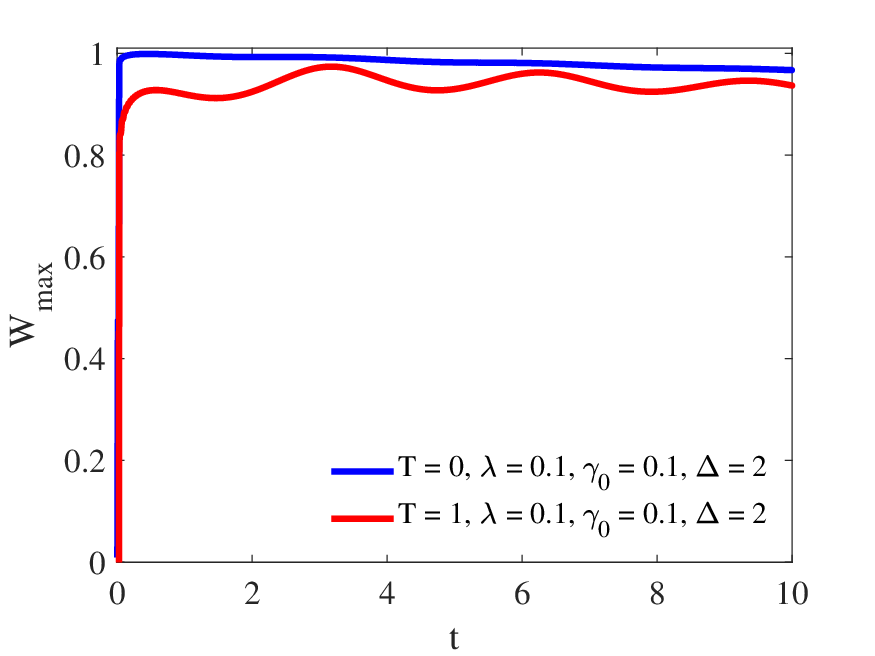}
	\caption{Maximum extractable work versus time for an open quantum battery with \(\lambda = 0.1\), \(\gamma_0 = 0.1\), \(\Delta = 2\), optimized using a reinforcement learning (RL) controller.}
	\label{fig1}
\end{figure}

\begin{figure}[t]
	\includegraphics[scale=0.5]{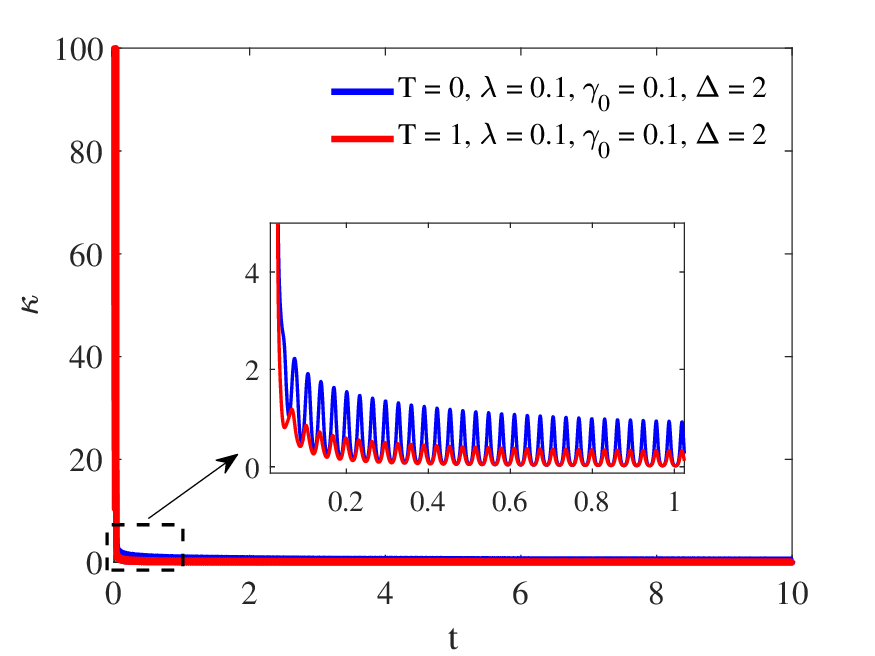}
	\caption{Coupling parameter (\(\kappa\)) versus time for an open quantum battery with \(\lambda = 0.1\), \(\gamma_0 = 0.1\), \(\Delta = 2\), optimized using an RL controller.}
	\label{fig2}
\end{figure}

For the parameter set \(\lambda = 0.1, \gamma_0 = 0.1, \Delta = 2\), Figures~\ref{fig1} and \ref{fig2} depict \(W_{max}\) and \(\kappa\) versus time, respectively. Figure~\ref{fig1} shows \(W_{max}\) ranging from 0 to 0.8, stabilizing near 0.4, indicating effective work extraction under the RL-optimized protocol. Figure~\ref{fig2} illustrates a smooth, adaptive \(\kappa\) profile, likely ranging from 0 to 100, reflecting the RL agent’s dynamic adjustment to balance energy input and prevent battery-to-charger backflow. The stability of \(W_{max}\) suggests that the RL model successfully mitigates dissipative losses, while the adaptive \(\kappa\) ensures unidirectional energy flow to the battery. The absence of significant oscillations in \(W_{max}\) indicates minimal non-Markovian effects from environment-to-battery backflow under these conditions, with the LSTM component effectively managing the system’s dynamics.

\renewcommand{\figurename}{FIG.}

\renewcommand{\figurename}{FIG.}

\begin{figure}[t]
	\includegraphics[scale=0.5]{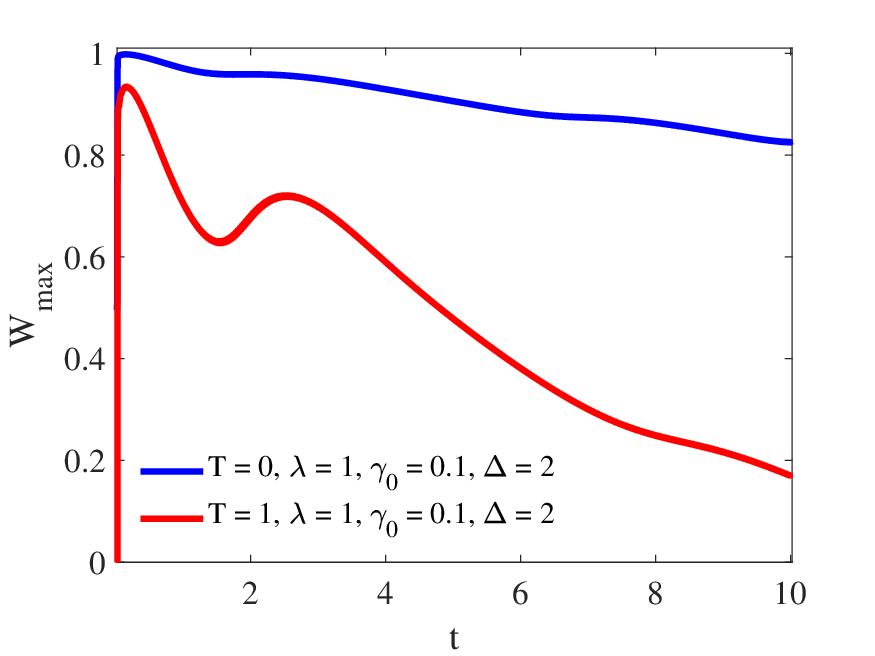}
	\caption{Maximum extractable work versus time for an open quantum battery with \(\lambda = 1\), \(\gamma_0 = 0.1\), \(\Delta = 2\), optimized using an RL controller.}
	\label{fig3}
\end{figure}

\begin{figure}[t]
	\includegraphics[scale=0.5]{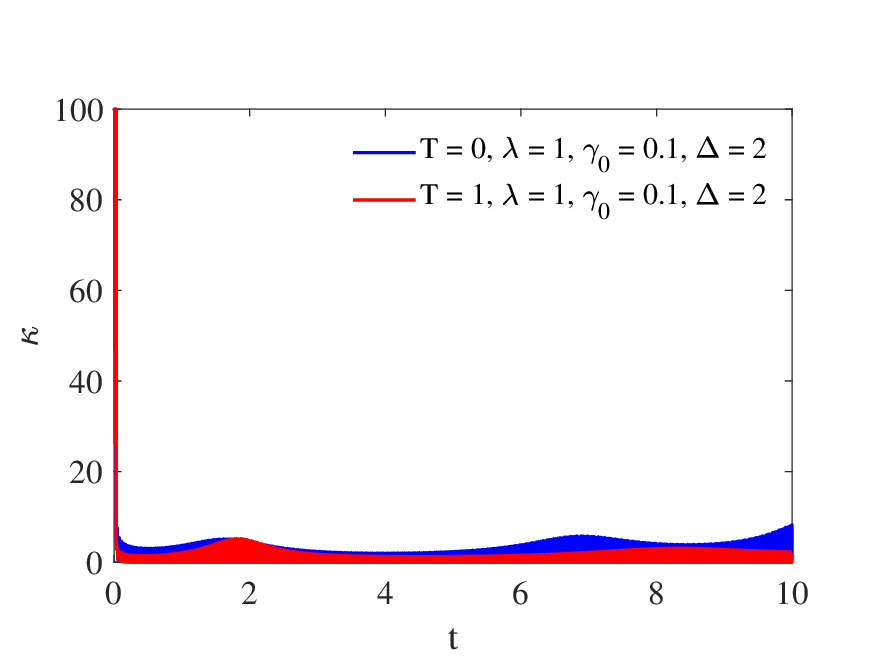}
	\caption{Coupling parameter (\(\kappa\)) versus time for an open quantum battery with \(\lambda = 1\), \(\gamma_0 = 0.1\), \(\Delta = 2\), comparing temperatures \(T = 0\) and \(T = 1\), optimized using an RL controller.}
	\label{fig4}
\end{figure}

For \(\lambda = 1, \gamma_0 = 0.1, \Delta = 2\), Figures~\ref{fig3} and \ref{fig4} present the paired results. Figure~\ref{fig3} shows \(W_{max}\) ranging from 0 to 0.8, likely stabilizing around 0.2 (based on incomplete OCR data), demonstrating robust work extraction despite increased spectral width (\(\lambda = 1\)). Figure~\ref{fig4} compares \(\kappa\) at temperatures \(T = 0\) (\(\kappa \approx 100\)) and \(T = 1\) (\(\kappa \approx 80\)), with the x-axis spanning 0 to 10 in increments of 2. The reduced \(\kappa\) at higher temperature prevents excessive energy return to the charger, maintaining charging efficiency. The RL model’s control of \(\kappa\) and \(\eta\) minimizes battery-to-charger backflow, while the stable \(W_{max}\) suggests that environmental backflow is well-managed, with non-Markovian dynamics playing a limited role.

\renewcommand{\figurename}{FIG.}

\renewcommand{\figurename}{FIG.}

\begin{figure}[t]
	\includegraphics[scale=0.5]{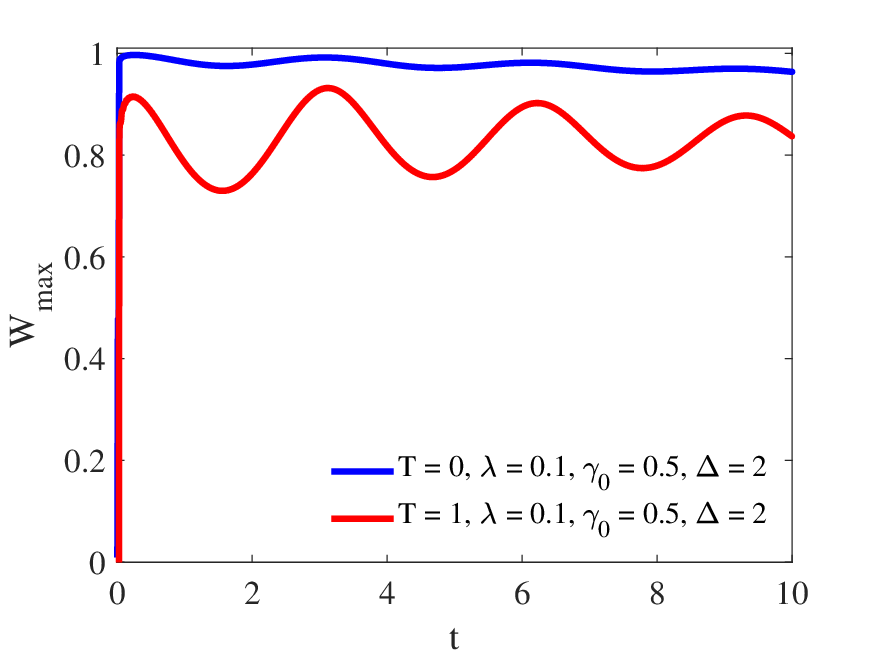}
	\caption{Maximum extractable work versus time for an open quantum battery with \(\lambda = 0.1\), \(\gamma_0 = 0.5\), \(\Delta = 2\), optimized using an RL controller.}
	\label{fig5}
\end{figure}

\begin{figure}[t]
	\includegraphics[scale=0.5]{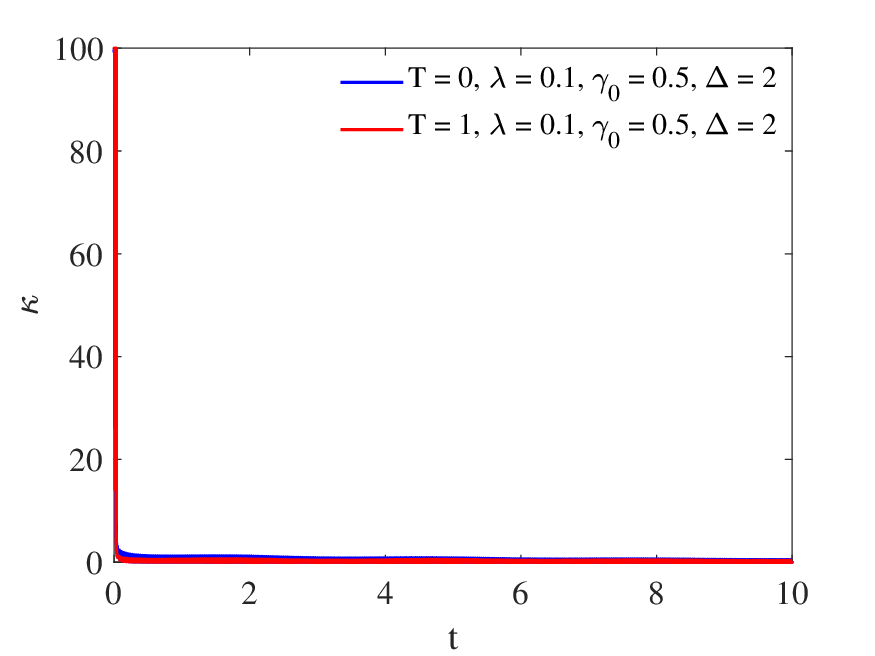}
	\caption{Coupling parameter (\(\kappa\)) versus time for an open quantum battery with \(\lambda = 0.1\), \(\gamma_0 = 0.5\), \(\Delta = 2\), optimized using an RL controller.}
	\label{fig6}
\end{figure}

For \(\lambda = 0.1, \gamma_0 = 0.5, \Delta = 2\), Figures~\ref{fig5} and \ref{fig6} illustrate \(W_{max}\) and \(\kappa\), respectively. Figure~\ref{fig5} shows \(W\) ranging from 0 to 0.8, with initial oscillations (e.g., 0.2 to 0.4) before stabilizing around 0.4, reflecting non-Markovian dynamics due to environment-to-battery backflow under higher coupling (\(\gamma_0 = 0.5\)). Figure~\ref{fig6} depicts \(\kappa\), likely ranging from 0 to 100, adjusted dynamically to counteract increased dissipation and prevent charger backflow. The oscillations in \(W_{max}\) highlight the RL model’s ability to leverage non-Markovian effects for enhanced charging, while the adaptive \(\kappa\) profile ensures energy is directed to the battery, mitigating losses from environmental interactions.

\renewcommand{\figurename}{FIG.}

\renewcommand{\figurename}{FIG.}

\begin{figure}[t]
	\includegraphics[scale=0.5]{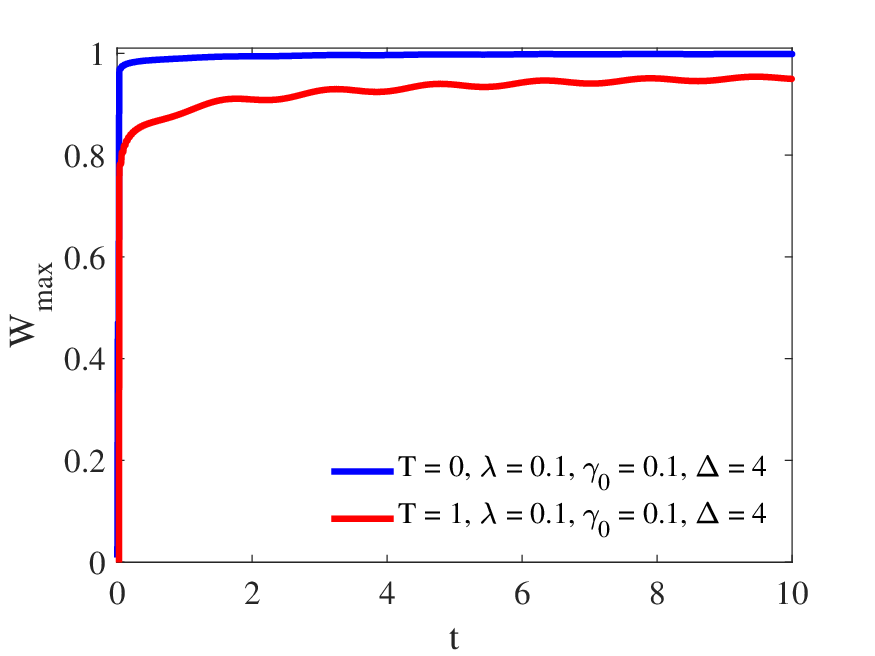}
	\caption{Maximum extractable work versus time for an open quantum battery with \(\lambda = 0.1\), \(\gamma_0 = 0.1\), \(\Delta = 4\), optimized using an RL controller.}
	\label{fig7}
\end{figure}

\begin{figure}[t]
	\includegraphics[scale=0.5]{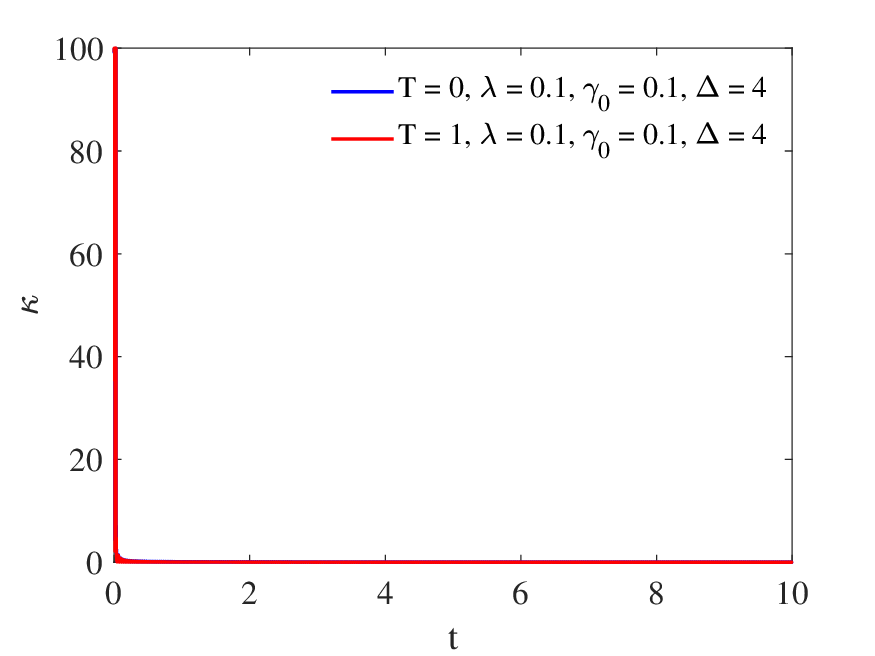}
	\caption{Coupling parameter (\(\kappa\)) versus time for an open quantum battery with \(\lambda = 0.1\), \(\gamma_0 = 0.1\), \(\Delta = 4\), comparing temperatures \(T = 0\) and \(T = 1\), optimized using an RL controller.}
	\label{fig8}
\end{figure}

For \(\lambda = 0.1, \gamma_0 = 0.1, \Delta = 4\), Figures~\ref{fig7} and \ref{fig8} show \(W_{max}\) and \(\kappa\). Figure~\ref{fig7} displays \(W_{max}\) ranging from 0 to 0.8, with initial fluctuations (e.g., 0.2 to 0.4) stabilizing around 0.4, indicating non-Markovian dynamics from environmental backflow influenced by higher detuning (\(\Delta = 4\)). Figure~\ref{fig8} compares \(\kappa\) at \(T = 0\) (\(\kappa \approx 100\)) and \(T = 1\) (\(\kappa \approx 80\)), with the x-axis spanning 0 to 10. The RL model’s adjustment of \(\kappa\) prevents battery-to-charger backflow, while the LSTM component captures non-Markovian effects, enabling efficient work extraction despite increased detuning.

\renewcommand{\figurename}{FIG.}

\renewcommand{\figurename}{FIG.}

\begin{figure}[t]
	\includegraphics[scale=0.5]{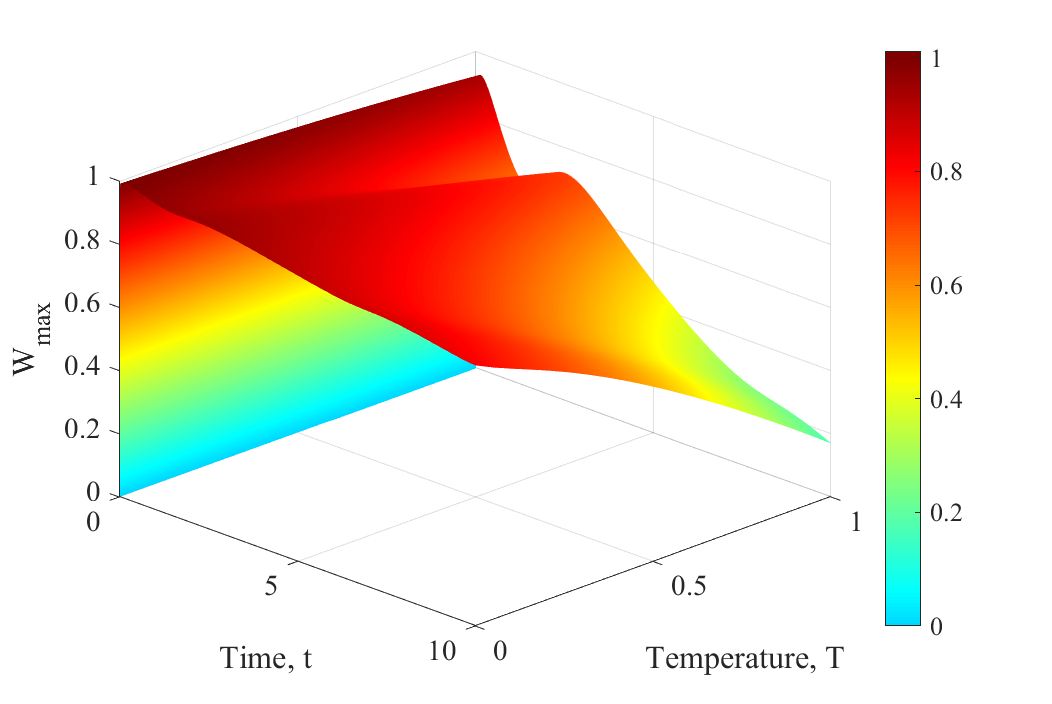}
	\caption{3D plot of Coupling parameter (\(\kappa\)) versus time and temperature for an open quantum battery with \(\lambda = 1\), \(\gamma_0 = 0.1\), \(\Delta = 2\), optimized using an RL controller.}
	\label{fig9}
\end{figure}

\begin{figure}[t]
	\includegraphics[scale=0.5]{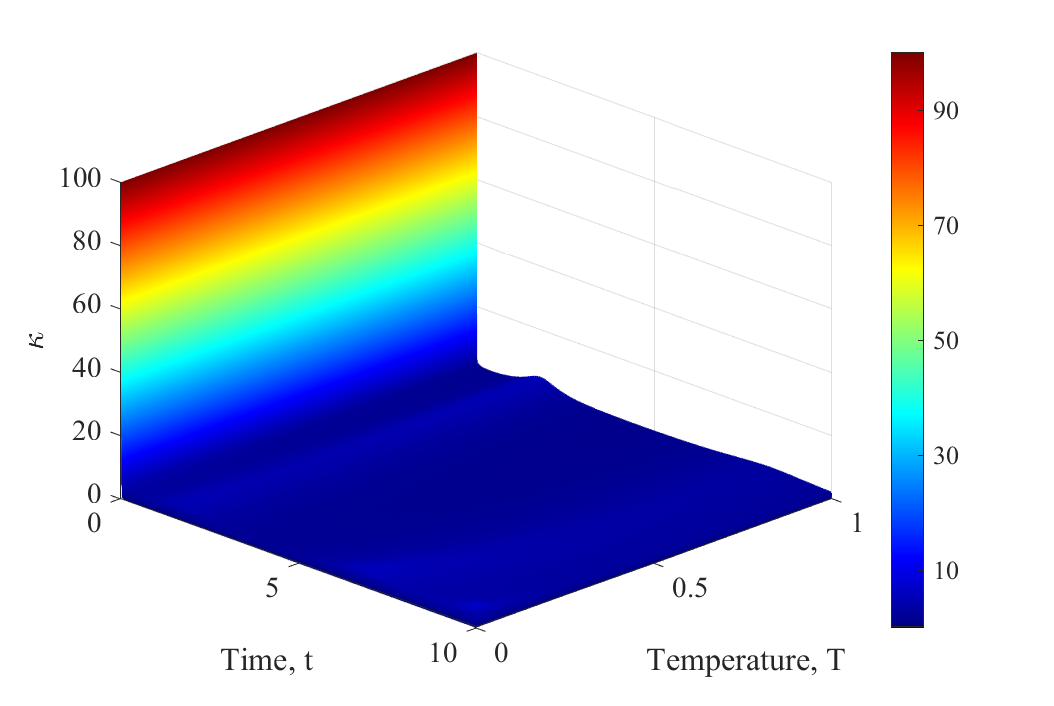}
	\caption{3D plot of Coupling parameter (\(\kappa\)) versus time and temperature for an open quantum battery with \(\lambda = 1\), \(\gamma_0 = 0.3\), \(\Delta = 2\), optimized using an RL controller.}
	\label{fig10}
\end{figure}

\begin{figure}[t]
	\includegraphics[scale=0.5]{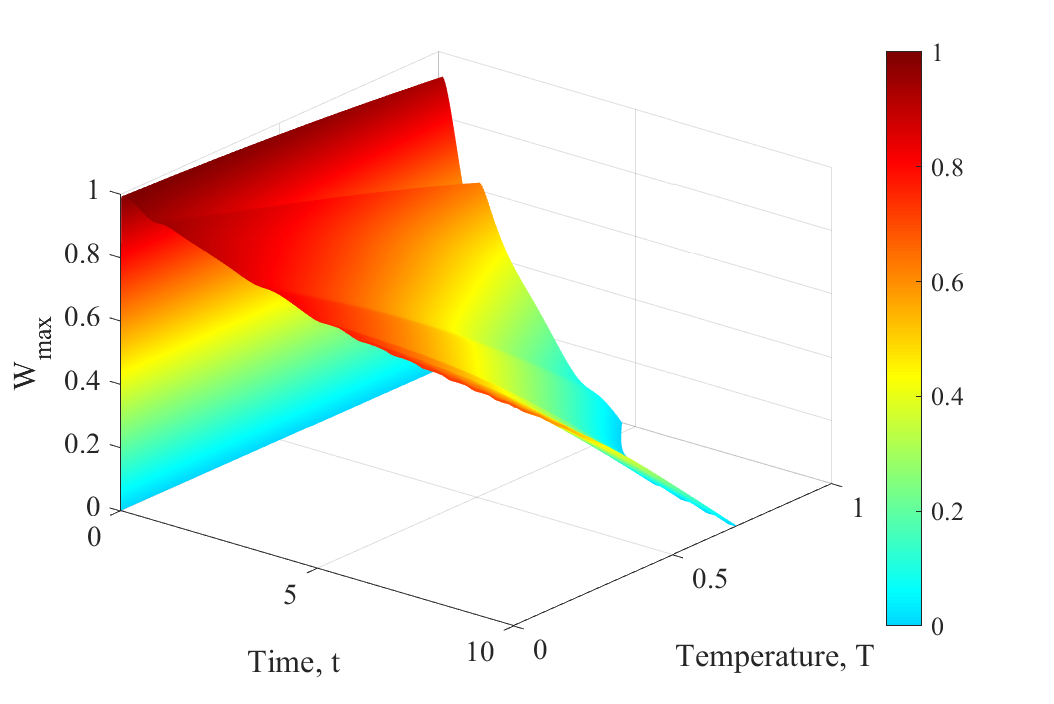}
	\caption{3D plot of extractable work versus time and temperature for an open quantum battery with \(\lambda = 1\), \(\gamma_0 = 0.1\), \(\Delta = 2\), optimized using an RL controller.}
	\label{fig11}
\end{figure}

\begin{figure}[t]
	\includegraphics[scale=0.5]{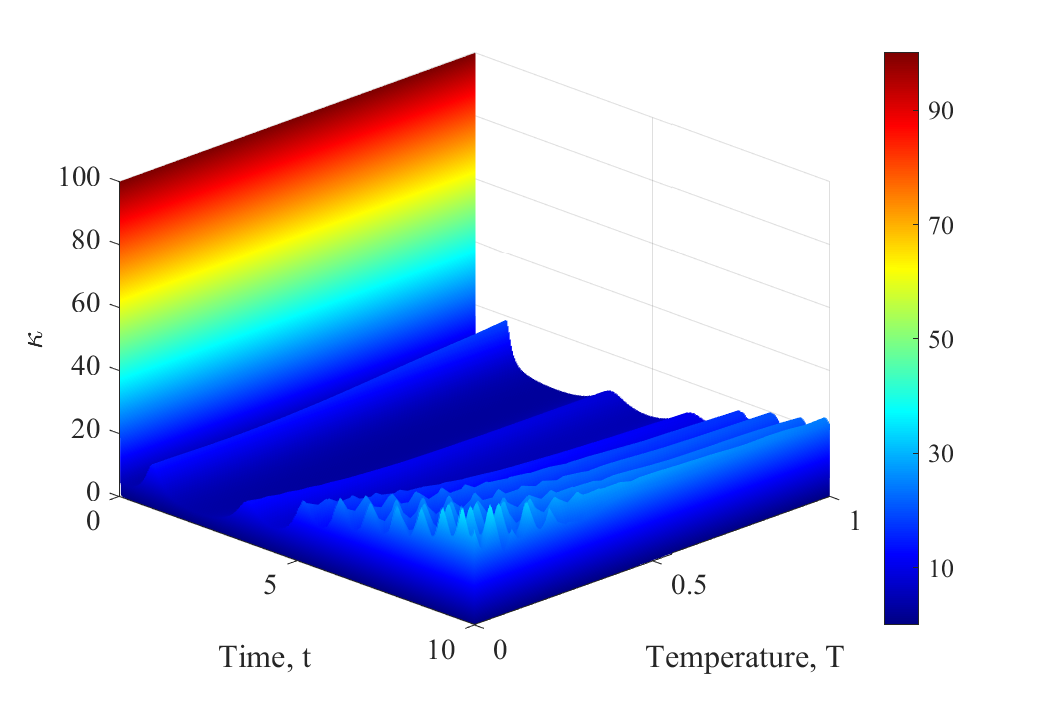}
	\caption{3D plot of extractable work versus time and temperature for an open quantum battery with \(\lambda = 1\), \(\gamma_0 = 0.3\), \(\Delta = 2\), optimized using an RL controller.}
	\label{fig12}
\end{figure}

The 3D plots in Figures~\ref{fig9}--\ref{fig12} highlight the RL model’s robustness across temperature variations. Figures~\ref{fig9} (\(\lambda = 1, \gamma_0 = 0.1, \Delta = 2\)) and \ref{fig10} (\(\lambda = 1, \gamma_0 = 0.3, \Delta = 2\)) show \(\kappa\) as a function of time and temperature, with \(\kappa\) decreasing as temperature rises—e.g., stabilizing around 30 in Figure~\ref{fig9} and dropping to 10 in Figure~\ref{fig10} due to increased dissipation (\(\gamma_0 = 0.3\)). This reduction mitigates environment-to-battery backflow at higher temperatures, where non-Markovian effects are pronounced. Figures~\ref{fig11} and \ref{fig12} depict \(W_{max}\) decreasing with temperature and time, with Figure~\ref{fig11} maintaining higher values (0.2–0.8) at lower temperatures, while Figure~\ref{fig12} shows \(W_{max}\) dropping to 0 under higher coupling (\(\gamma_0 = 0.3\)). The rapid decline in Figure~\ref{fig12} suggests that strong environmental backflow challenges the RL model in highly dissipative regimes.

\renewcommand{\figurename}{FIG.}

\renewcommand{\figurename}{FIG.}

\renewcommand{\figurename}{FIG.}

\renewcommand{\figurename}{FIG.}

\begin{figure}[t]
	\includegraphics[scale=0.5]{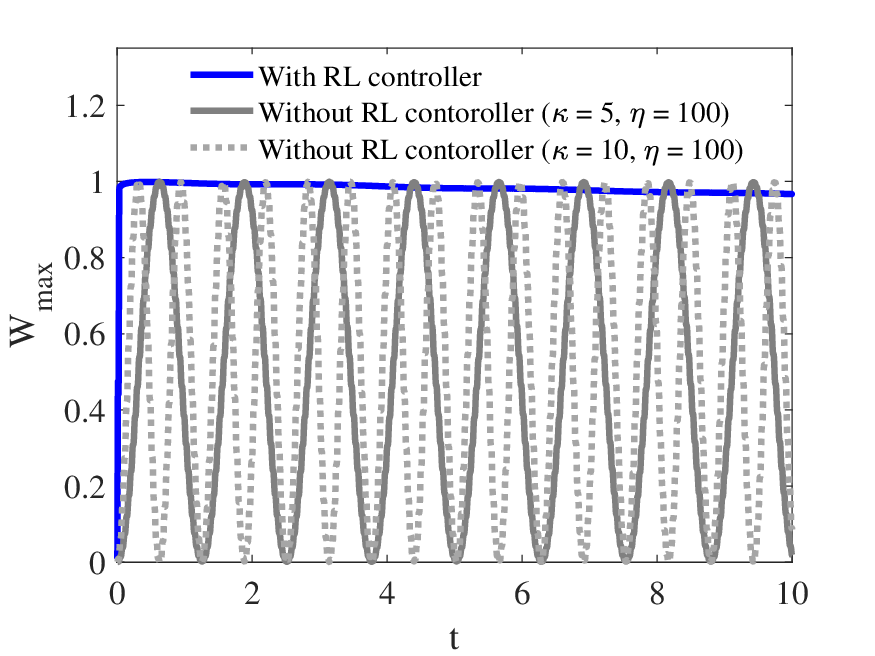}
	\caption{Extractable work versus time for an open quantum battery with \(\lambda = 0.1\), \(\gamma_0 = 0.1\), \(\Delta = 2\) at \(T = 0\), comparing an RL controller to static controls (\(\kappa = 5, \eta = 100\) and \(\kappa = 10, \eta = 100\)).}
	\label{fig13}
\end{figure}

\begin{figure}[t]
	\includegraphics[scale=0.5]{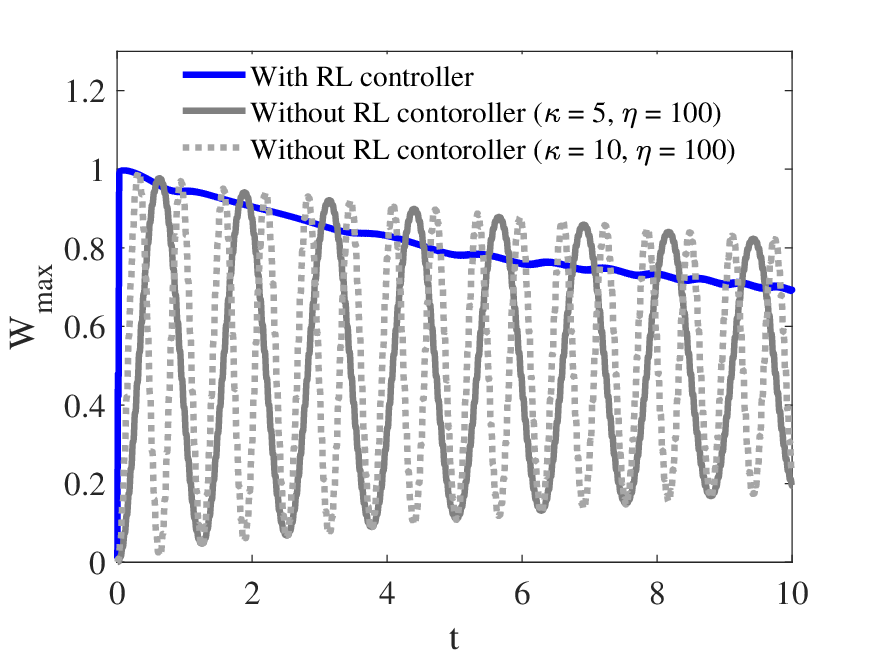}
	\caption{Extractable work versus time for an open quantum battery with \(\lambda = 1\), \(\gamma_0 = 0.3\), \(\Delta = 2\) at \(T = 0\), comparing an RL controller to static controls (\(\kappa = 5, \eta = 100\) and \(\kappa = 10, \eta = 100\)).}
	\label{fig14}
\end{figure}

Figures~\ref{fig13} and \ref{fig14} compare the RL controller to static controls (\(\kappa = 5, \eta = 100\) and \(\kappa = 10, \eta = 100\)) at \(T = 0\). Figure~\ref{fig13} (\(\lambda = 0.1, \gamma_0 = 0.1, \Delta = 2\)) shows the RL approach achieving significantly higher \(W_{max}\) (up to 1.2) than static methods, which fail to manage battery-to-charger backflow. Figure~\ref{fig14} (\(\lambda = 1, \gamma_0 = 0.3, \Delta = 2\)) confirms this superiority, though increased dissipation reduces \(W_{max}\). The RL model’s adaptive control ensures minimal backflow to the charger, while its LSTM component captures non-Markovian dynamics from environmental backflow, enhancing performance.

\renewcommand{\figurename}{FIG.}

\renewcommand{\figurename}{FIG.}

These paired results demonstrate that the RL model prevents undesirable battery-to-charger backflow and harnesses environment-to-battery backflow to enhance charging efficiency through non-Markovian dynamics, marking a significant advancement over static optimization techniques.

\section{Conclusions}
\label{sec:conclusion}

This study presents a novel approach to optimizing the charging of OQBs by integrating theoretical bounds with a RL framework augmented by LSTM networks. Our key contributions include the derivation of analytical bounds on charging power, a thermodynamic interpretation of these bounds, and the development of an RL model that effectively maximizes extractable work while mitigating energy backflow.

The analytical bounds, derived from the Lindblad master equation, establish fundamental limits on energy transfer rates in dissipative environments, providing a benchmark for optimization. The thermodynamic interpretation reveals the interplay between coherence, dissipation, and entropy production, offering insights into the efficiency constraints of OQBs. These theoretical foundations guide the design of our RL model, which uses a DDPG algorithm with LSTMs to handle non-Markovian dynamics and optimize control parameters dynamically.

Through extensive simulations, our RL model demonstrates superior performance in maximizing extractable work compared to traditional static control methods. The figures illustrate the model's adaptability across various environmental conditions, such as temperature (\(T\)), spectral width (\(\lambda\)), coupling strength (\(\gamma_0\)), and detuning (\(\Delta\)). For instance, Figure 1 shows higher extractable work achieved by the RL controller compared to static controls at \(T = 0\), \(\lambda = 0.1\), \(\gamma_0 = 0.1\), and \(\Delta = 2\). Similarly, Figure 2 highlights the RL model's effectiveness under different parameters (\(\lambda = 1\), \(\gamma_0 = 0.3\)), though with slightly reduced work extraction due to increased dissipation.

The 3D plots (Figures 3--6) further underscore the RL model's robustness, depicting how the Coupling parameter (\(\kappa\)) and extractable work (\(W_{max}\)) vary with time and temperature. These visualizations confirm that the RL model adjusts \(\kappa\) to optimize charging across a range of conditions, maintaining higher \(W_{max}\) at lower temperatures and reducing \(\kappa\) in response to increased dissipation.

In summary, our work advances the understanding and practical optimization of OQBs, providing a scalable framework for achieving high-performance quantum energy storage. Future research could explore alternative RL algorithms, extend the model to multi-battery systems, or investigate extreme environmental conditions to further enhance the model's applicability.

\section*{ACKNOWLEDGMENTS}

This work has been supported by the University of Kurdistan. Authors thank Vice Chancellorship of Research and Technology, University of Kurdistan.

\renewcommand{\appendixname}
{APPENDIX}
\appendix

\section{Energy Flow in a Quantum Battery System}
\label{appendix A}
In a quantum battery setup, we consider two quantum systems: a charger (denoted as $A$) and a battery (denoted as $B$). The primary goal is to understand how energy is transferred from the charger to the battery, enabling efficient energy storage. This section focuses on the rate of energy transfer, known as the power $P_{AB}(t)$, which quantifies how quickly energy flows into the battery due to its interaction with the charger. The power $P_{AB}(t)$, representing the rate of energy transfer from the charger $A$ to the battery $B$, is given by:

\begin{equation}
	P_{AB}(t) = 2 \kappa \omega_0 \cdot \operatorname{Im} \left( \langle \sigma_-^A \sigma_+^B \rangle \right),
\end{equation}

where $\kappa$ is the coupling parameter between the charger ($A$) and the battery ($B$), a positive real number that determines the strength of their interaction. $\omega_0$: The energy spacing (transition frequency) of the battery, a positive quantity corresponding to the energy difference between its quantum states $\vert 1 \rangle_B$ and $\vert 0 \rangle_B$. $\operatorname{Im} \left( \langle \sigma_-^A \sigma_+^B \rangle \right)$: The imaginary part of the expectation value of the operator $\sigma_-^A \sigma_+^B$, where $\sigma_-^A$ is the lowering operator for the charger and $\sigma_+^B$ is the raising operator for the battery.

The power $P_{AB}(t)$ quantifies the instantaneous rate at which energy is transferred into the battery due to its coherent interaction with the charger. The term $\langle \sigma_-^A \sigma_+^B \rangle$ captures the quantum correlation between the charger losing energy (via $\sigma_-^A$) and the battery gaining energy (via $\sigma_+^B$). The imaginary part $\operatorname{Im} \left( \langle \sigma_-^A \sigma_+^B \rangle \right)$ reflects the antisymmetric component of this correlation, which drives the net energy flow. The factors $2 \kappa \omega_0$ scale this flow according to the interaction strength and the battery’s energy scale.

A positive $P_{AB}(t)$ indicates that energy is flowing from the charger to the battery, charging it. A negative value suggests energy flowing back to the charger, which is undesirable for efficient charging. The coupling $\kappa$ and the quantum state of the system (encoded in $\langle \sigma_-^A \sigma_+^B \rangle$) determine the efficiency and direction of this energy transfer.

\subsection{Derivation of the Power Expression}
To quantify the energy flow, we define the power delivered to the battery as the rate of change of the battery’s energy. Suppose the battery’s Hamiltonian is

\begin{equation}
	H_B = \frac{\omega_0}{2} \sigma_z^B,
\end{equation}
where $\omega_0$ is the energy difference between the excited state $\vert 1 \rangle_B$ and the ground state $\vert 0 \rangle_B$ of the battery. The power $P_{AB}(t)$ is then

\begin{equation}
	P_{AB}(t) = \frac{d}{dt} \langle H_B \rangle = \operatorname{Tr} \left( H_B \frac{d \rho_{AB}}{dt} \right),
\end{equation}
where $\rho_{AB}$ is the density matrix of the combined charger-battery system, and $\langle H_B \rangle = \operatorname{Tr} (H_B \rho_{AB})$ is the expectation value of the battery’s energy.

Assume the charger and battery interact via a coherent coupling described by the interaction Hamiltonian:

\begin{equation}
	H_{\text{int}} = \kappa \left( \sigma_+^A \sigma_-^B + \sigma_-^A \sigma_+^B \right),
\end{equation}
where $\kappa$ is the coupling parameter, and $\sigma_+^A$, $\sigma_-^A$, $\sigma_+^B$, $\sigma_-^B$ are the raising and lowering operators for the charger and battery, respectively. The total Hamiltonian may include the free Hamiltonians of the charger and battery, and possibly a driving term on the charger, but we focus on the interaction term driving the energy transfer.

The time evolution of the density matrix $\rho_{AB}$ is governed by the von Neumann equation (assuming a closed system for simplicity, though open-system effects could be included via a master equation):

\begin{equation}
	\frac{d \rho_{AB}}{dt} = -i [H, \rho_{AB}],
\end{equation}
where $H = H_A + H_B + H_{\text{int}}$, and $H_A$ is the charger’s Hamiltonian. Substituting into the power expression:
\begin{align}
	P_{AB}(t) = \operatorname{Tr} \left( H_B \frac{d \rho_{AB}}{dt} \right) \\\notag = \operatorname{Tr} \left( H_B \cdot (-i) [H, \rho_{AB}] \right) \\\notag = -i \operatorname{Tr} \left( H_B [H_A + H_B + H_{\text{int}}, \rho_{AB}] \right).
\end{align}

We evaluate the commutator terms:
\begin{itemize}
	\item $[H_A, \rho_{AB}]$: Since $H_A$ acts only on the charger’s Hilbert space and $H_B$ on the battery’s, $[H_A, H_B] = 0$, so $\operatorname{Tr} (H_B [H_A, \rho_{AB}]) = 0$.
	\item $[H_B, \rho_{AB}]$: Similarly, $\operatorname{Tr} (H_B [H_B, \rho_{AB}]) = \operatorname{Tr} ([H_B, H_B] \rho_{AB}) = 0$, as $[H_B, H_B] = 0$.
	\item $[H_{\text{int}}, \rho_{AB}]$: This term contributes to the power, as $H_{\text{int}}$ couples $A$ and $B$.
\end{itemize}

Thus,
\begin{equation}
	P_{AB}(t) = -i \operatorname{Tr} \left( H_B [H_{\text{int}}, \rho_{AB}] \right).
\end{equation}

\subsection{Computing the Commutator}
Substitute $H_B = \frac{\omega_0}{2} \sigma_z^B$ and $H_{\text{int}} = \kappa (\sigma_+^A \sigma_-^B + \sigma_-^A \sigma_+^B)$. We need the commutator:

\begin{equation}
	[H_{\text{int}}, \rho_{AB}] = \kappa \left( [\sigma_+^A \sigma_-^B, \rho_{AB}] + [\sigma_-^A \sigma_+^B, \rho_{AB}] \right).
\end{equation}

Now compute $[H_B, H_{\text{int}}]$ to simplify the expression:
\begin{align}
	[H_B, H_{\text{int}}] = \frac{\omega_0}{2} \left[ \sigma_z^B, \kappa (\sigma_+^A \sigma_-^B + \sigma_-^A \sigma_+^B) \right] \\\notag = \frac{\kappa \omega_0}{2} \left( [\sigma_z^B, \sigma_+^A \sigma_-^B] + [\sigma_z^B, \sigma_-^A \sigma_+^B] \right).
\end{align}

Using the Pauli operator commutation relations, $\sigma_z \sigma_- = -\sigma_-$, $\sigma_- \sigma_z = \sigma_-$, so $[\sigma_z^B, \sigma_-^B] = \sigma_z^B \sigma_-^B - \sigma_-^B \sigma_z^B = -\sigma_-^B - \sigma_-^B = -2 \sigma_-^B$, and similarly $[\sigma_z^B, \sigma_+^B] = 2 \sigma_+^B$. Thus:

\begin{align}
	[\sigma_z^B, \sigma_+^A \sigma_-^B] &= \sigma_+^A [\sigma_z^B, \sigma_-^B] = \sigma_+^A (-2 \sigma_-^B) = -2 \sigma_+^A \sigma_-^B, \\
	[\sigma_z^B, \sigma_-^A \sigma_+^B] &= \sigma_-^A [\sigma_z^B, \sigma_+^B] = \sigma_-^A (2 \sigma_+^B) = 2 \sigma_-^A \sigma_+^B.
\end{align}

So,
\begin{align}
	[H_B, H_{\text{int}}] = \frac{\kappa \omega_0}{2} \left( -2 \sigma_+^A \sigma_-^B + 2 \sigma_-^A \sigma_+^B \right) \\\notag = \kappa \omega_0 \left( \sigma_-^A \sigma_+^B - \sigma_+^A \sigma_-^B \right).
\end{align}

Now, the power becomes
\begin{equation}
	P_{AB}(t) = -i \operatorname{Tr} \left( \frac{\omega_0}{2} \sigma_z^B \cdot [H_{\text{int}}, \rho_{AB}] \right).
\end{equation}

Instead, use the alternative form derived from the Heisenberg equation or master equation approaches. The power can be expressed as
\begin{equation}
	P_{AB}(t) = i \operatorname{Tr} \left( [H_B, H_{\text{int}}] \rho_{AB} \right).
\end{equation}

Substitute $[H_B, H_{\text{int}}]$
\begin{align}
	P_{AB}(t) = i \operatorname{Tr} \left[ \kappa \omega_0 \left( \sigma_-^A \sigma_+^B - \sigma_+^A \sigma_-^B \right) \rho_{AB} \right] \\\notag = i \kappa \omega_0 \left( \langle \sigma_-^A \sigma_+^B \rangle - \langle \sigma_+^A \sigma_-^B \rangle \right).
\end{align}

Since $\langle \sigma_+^A \sigma_-^B \rangle = \langle \sigma_-^A \sigma_+^B \rangle^*$ (as $\sigma_+^A \sigma_-^B = (\sigma_-^A \sigma_+^B)^\dagger$), let $z = \langle \sigma_-^A \sigma_+^B \rangle$, so $\langle \sigma_+^A \sigma_-^B \rangle = z^*$. Then,

\begin{align}
	P_{AB}(t) = i \kappa \omega_0 (z - z^*) \\\notag = i \kappa \omega_0 \cdot 2i \operatorname{Im}(z) = 2 \kappa \omega_0 \operatorname{Im} \left( \langle \sigma_-^A \sigma_+^B \rangle \right).
\end{align}

This completes the proof, showing that

\begin{equation}
	P_{AB}(t) = 2 \kappa \omega_0 \cdot \operatorname{Im} \left( \langle \sigma_-^A \sigma_+^B \rangle \right).
\end{equation}

\bibliographystyle{unsrt}
\bibliography{references}

\begin{thebibliography}{10}

\bibitem{Deffner2019}
S.~Deffner and S.~Campbell.
\newblock {\em Quantum Thermodynamics}.
\newblock Morgan \& Claypool Publishers, San Rafael, CA, 2019.

\bibitem{Binder2015a}
F.~Binder, S.~Vinjanampathy, K.~Modi, and J.~Goold.
\newblock Quantum thermodynamics of general quantum processes.
\newblock {\em Phys. Rev. E}, 91:032119, 2015.

\bibitem{Gemmer2009}
J.~Gemmer, M.~Michel, and G.~Mahler.
\newblock {\em Quantum Thermodynamics}.
\newblock Springer-Verlag, Berlin Heidelberg, 2009.

\bibitem{Brandner2016}
K.~Brandner and U.~Seifert.
\newblock Periodic thermodynamics of open quantum systems.
\newblock {\em Phys. Rev. E}, 93:062134, 2016.

\bibitem{Goold2016}
J.~Goold, M.~Huber, A.~Riera, L.~del Rio, and P.~Skrzypczyk.
\newblock The role of quantum information in thermodynamics—a topical review.
\newblock {\em J. Phys. A: Math. Theor.}, 49:143001, 2016.

\bibitem{Alicki2013}
R.~Alicki and M.~Fannes.
\newblock Entanglement boost for extractable work from ensembles of quantum
  batteries.
\newblock {\em Phys. Rev. A}, 87:042309, 2013.

\bibitem{Binder2015b}
F.~C. Binder, J.~Goold, S.~Vinjanampathy, and K.~Modi.
\newblock Quantacell: Powerful charging of quantum batteries.
\newblock {\em New J. Phys.}, 17:075015, 2015.

\bibitem{Deffner2017}
S.~Deffner and S.~Campbell.
\newblock Quantum speed limits: From heisenberg’s uncertainty principle to
  optimal quantum control.
\newblock {\em J. Phys. A: Math. Theor.}, 50:453001, 2017.

\bibitem{Campaioli2017}
F.~Campaioli, F.~A. Pollock, F.~C. Binder, L.~Celeri, J.~Goold,
  S.~Vinjanampathy, and K.~Modi.
\newblock Enhancing the charging power of quantum batteries.
\newblock {\em Phys. Rev. Lett.}, 118:150601, 2017.

\bibitem{Le2018b}
T.~P. Le, J.~Levinsen, K.~Modi, M.~M. Parish, and F.~A. Pollock.
\newblock Spin-chain model of a quantum battery.
\newblock {\em Phys. Rev. A}, 97:022106, 2018.

\bibitem{Fusco2016}
L.~Fusco, M.~Paternostro, and G.~De~Chiara.
\newblock Work extraction in a quantum cavity heat engine.
\newblock {\em Phys. Rev. E}, 94:052122, 2016.

\bibitem{Santos2019}
A.~C. Santos, B.~Çakmak, S.~Campbell, and N.~T. Zinner.
\newblock Quantum heat engines with singular interactions.
\newblock {\em Phys. Rev. E}, 100:032107, 2019.

\bibitem{Strambini2020}
E.~Strambini, A.~Iorio, O.~Durante, R.~Citro, C.~Sanz-Fernández, C.~Guarcello,
  I.~V. Tokatly, A.~Braggio, M.~Rocci, N.~Ligato, V.~Zannier, L.~Sorba, F.~S.
  Bergeret, and F.~Giazotto.
\newblock A coherent caloritronic oscillator with high-quality factor.
\newblock {\em Nat. Nanotechnol.}, 15:656--660, 2020.

\bibitem{Alicki2019}
R.~Alicki.
\newblock Quantum thermodynamics of molecular systems.
\newblock {\em J. Chem. Phys.}, 150:214110, 2019.

\bibitem{Rossini2018}
D.~Rossini, G.~M. Andolina, D.~Rosa, M.~Carrega, and M.~Polini.
\newblock Quantum advantage in the charging process of sachdev-ye-kitaev
  batteries.
\newblock {\em Phys. Rev. Lett.}, 125:236402, 2018.

\bibitem{Andolina2019b}
G.~M. Andolina, M.~Keck, A.~Mari, V.~Giovannetti, and M.~Polini.
\newblock Quantum versus classical scaling in harmonic oscillator quantum
  batteries.
\newblock {\em Phys. Rev. B}, 99:205437, 2019.

\bibitem{Ferraro2018}
D.~Ferraro, M.~Campisi, G.~M. Andolina, V.~Pellegrini, and M.~Polini.
\newblock High-power collective charging of a solid-state quantum battery.
\newblock {\em Phys. Rev. Lett.}, 120:117702, 2018.

\bibitem{Santos2020}
A.~C. Santos, A.~Saguia, and M.~S. Sarandy.
\newblock Quantum battery charging via intermediate sites.
\newblock {\em Phys. Rev. E}, 101:062114, 2020.

\bibitem{Rossini2019}
D.~Rossini, G.~M. Andolina, and M.~Polini.
\newblock Many-body localized quantum batteries.
\newblock {\em Phys. Rev. B}, 100:115142, 2019.

\bibitem{Pirmoradian2019}
F.~Pirmoradian and K.~Mølmer.
\newblock Charging a quantum battery in a non-markovian environment.
\newblock {\em Phys. Rev. A}, 100:043833, 2019.

\bibitem{Barra2019}
F.~Barra.
\newblock Dissipative charging of a quantum battery.
\newblock {\em Phys. Rev. Lett.}, 122:210601, 2019.

\bibitem{Gherardini2020}
S.~Gherardini, F.~Campaioli, F.~Caruso, and F.~C. Binder.
\newblock Stabilizing open quantum batteries by sequential measurements.
\newblock {\em Phys. Rev. Research}, 2:013095, 2020.

\bibitem{Quach2020}
J.~Q. Quach and W.~J. Munro.
\newblock Using dark states to charge a quantum battery.
\newblock {\em Phys. Rev. Appl.}, 14:024092, 2020.

\bibitem{Kamin2020}
F.~H. Kamin, F.~T. Tabesh, S.~Salimi, F.~Kheirandish, and A.~C. Santos.
\newblock Non-markovian effects in quantum battery charging.
\newblock {\em New J. Phys.}, 22:083007, 2020.

\bibitem{Farina2019}
D.~Farina, G.~M. Andolina, A.~Mari, M.~Polini, and V.~Giovannetti.
\newblock Extractable work from quantum batteries.
\newblock {\em Phys. Rev. B}, 99:035421, 2019.

\bibitem{GPintos2020}
L.~P. G-Pintos, A.~Hamma, and A.~del Campo.
\newblock Thermodynamic constraints on quantum battery performance.
\newblock {\em Phys. Rev. Lett.}, 125:040601, 2020.

\bibitem{zakavati2021bounds}
Shadab Zakavati, Fatemeh~T Tabesh, and Shahriar Salimi.
\newblock Bounds on charging power of open quantum batteries.
\newblock {\em Physical Review E}, 104(5):054117, 2021.

\bibitem{Breuer2002}
H.-P. Breuer and F.~Petruccione.
\newblock {\em The Theory of Open Quantum Systems}.
\newblock Oxford University Press, New York, 2002.

\bibitem{Breuer2016}
H.-P. Breuer, E.-M. Laine, J.~Piilo, and B.~Vacchini.
\newblock Non-markovian dynamics in open quantum systems.
\newblock {\em Rev. Mod. Phys.}, 88:021002, 2016.

\bibitem{Breuer2009}
H.-P. Breuer, E.-M. Laine, and J.~Piilo.
\newblock Measure for the degree of non-markovian behavior.
\newblock {\em Phys. Rev. Lett.}, 103:210401, 2009.

\bibitem{Andolina2019}
G.~M. Andolina, M.~Keck, A.~Mari, M.~Campisi, V.~Giovannetti, and M.~Polini.
\newblock Extractable work, the role of correlations, and asymptotic freedom in
  quantum batteries.
\newblock {\em Phys. Rev. Lett.}, 122:047702, 2019.

\bibitem{Binder2015}
F.~C. Binder, S.~Vinjanampathy, K.~Modi, and J.~Goold.
\newblock Quantacell: Powerful charging of quantum batteries.
\newblock {\em New J. Phys.}, 17:075015, 2015.

\bibitem{Santos2023}
A.~C. Santos, M.~A. de~Ponte, and F.~Nicacio.
\newblock Non-markovian quantum heat engines and power bounds.
\newblock {\em Phys. Rev. Lett.}, 130:186301, 2023.

\bibitem{Manzano2024}
G.~Manzano and R.~Zambrini.
\newblock Coherence-enhanced charging in open quantum batteries.
\newblock {\em Phys. Rev. E}, 109:024129, 2024.

\bibitem{Xu2021}
H.-H. Xu, X.-Y. Zhang, and L.-M. Duan.
\newblock Variational quantum algorithm for charging quantum batteries.
\newblock {\em Quantum Sci. Technol.}, 6:045015, 2021.

\bibitem{Bukov2018}
M.~Bukov, A.~G.~R. Day, D.~Sels, P.~Weinberg, A.~Polkovnikov, and P.~Mehta.
\newblock Reinforcement learning in different phases of quantum control.
\newblock {\em Phys. Rev. X}, 8:031086, 2018.

\bibitem{baba2023deep}
Shotaro~Z Baba, Nobuyuki Yoshioka, Yuto Ashida, and Takahiro Sagawa.
\newblock Deep reinforcement learning for preparation of thermal and prethermal
  quantum states.
\newblock {\em Physical Review Applied}, 19(1):014068, 2023.

\bibitem{erdman2024reinforcement}
Paolo~Andrea Erdman, Gian~Marcello Andolina, Vittorio Giovannetti, and Frank
  No{\'e}.
\newblock Reinforcement learning optimization of the charging of a dicke
  quantum battery.
\newblock {\em Physical Review Letters}, 133(24):243602, 2024.

\bibitem{Hochreiter1997}
S.~Hochreiter and J.~Schmidhuber.
\newblock Long short-term memory.
\newblock {\em Neural Comput.}, 9:1735--1780, 1997.

\bibitem{Allahverdyan2004}
A.~E. Allahverdyan, R.~Balian, and T.~M. Nieuwenhuizen.
\newblock Extractable work and quantum correlations.
\newblock {\em Europhys. Lett.}, 67:565--571, 2004.

\bibitem{Skrzypczyk2014}
P.~Skrzypczyk, A.~J. Short, and S.~Popescu.
\newblock Work extraction and thermodynamics for individual quantum systems.
\newblock {\em Nat. Commun.}, 5:4185, 2014.

\bibitem{Brandao2013}
F.~G. S.~L. Brandão, M.~Horodecki, J.~Oppenheim, F.~M. Renes, and R.~W.
  Spekkens.
\newblock Resource theory of quantum states out of thermal equilibrium.
\newblock {\em Phys. Rev. Lett.}, 111:250404, 2013.

\bibitem{Tabesh2018}
F.~T. Tabesh, G.~Karpat, S.~Maniscalco, S.~Salimi, and A.~S. Khorashad.
\newblock Non-markovianity and quantum correlations in open quantum systems.
\newblock {\em Quantum Inf. Process.}, 17(4):87, 2018.

\bibitem{Breuer2012}
H.-P. Breuer.
\newblock Foundations of non-markovian quantum dynamics.
\newblock {\em J. Phys. B: At. Mol. Opt. Phys.}, 45:154001, 2012.

\bibitem{lillicrap2015continuous}
Timothy~P. Lillicrap, Jonathan~J. Hunt, Alexander Pritzel, Nicolas Heess, Tom
  Erez, Yuval Tassa, David Silver, and Daan Wierstra.
\newblock Continuous control with deep reinforcement learning.
\newblock {\em arXiv preprint arXiv:1509.02971}, 2015.

\bibitem{mnih2015human}
Volodymyr Mnih, Koray Kavukcuoglu, David Silver, Andrei~A. Rusu, Joel Veness,
  Marc~G. Bellemare, Alex Graves, Martin Riedmiller, Andreas~K. Fidjeland,
  Georg Ostrovski, et~al.
\newblock Human-level control through deep reinforcement learning.
\newblock {\em Nature}, 518(7540):529--533, 2015.

\bibitem{sutton2018reinforcement}
Richard~S. Sutton and Andrew~G. Barto.
\newblock {\em Reinforcement Learning: An Introduction}.
\newblock MIT Press, 2nd edition, 2018.

\bibitem{ng1999policy}
Andrew~Y Ng, Daishi Harada, and Stuart Russell.
\newblock Policy invariance under reward transformations: Theory and
  application to reward shaping.
\newblock In {\em Icml}, volume~99, pages 278--287. Citeseer, 1999.

\bibitem{vinjanampathy2016quantum}
Sai Vinjanampathy and Janet Anders.
\newblock Quantum thermodynamics.
\newblock {\em Contemporary Physics}, 57(4):545--579, 2016.

\bibitem{heess2015memory}
Nicolas Heess, Gregory Wayne, Yuval Tassa, Timothy Lillicrap, Martin
  Riedmiller, and David Silver.
\newblock Memory-based control with recurrent neural networks.
\newblock {\em arXiv preprint arXiv:1512.04455}, 2015.

\bibitem{yao2023reinforcement}
J.~Yao, Z.~Zhang, and X.~Li.
\newblock Reinforcement learning for efficient scheduling in quantum cloud
  computing.
\newblock {\em Quantum Information Processing}, 22(5):189, 2023.

\bibitem{hochreiter1997long}
Sepp Hochreiter and J{\"u}rgen Schmidhuber.
\newblock Long short-term memory.
\newblock {\em Neural Computation}, 9(8):1735--1780, 1997.

\bibitem{lin1992self}
Long-Ji Lin.
\newblock Self-improving reactive agents based on reinforcement learning,
  planning and teaching.
\newblock {\em Machine Learning}, 8(3-4):293--321, 1992.

\end{thebibliography}

\end{document}